# Deep manifold learning reveals hidden dynamics of proteasome autoregulation

Zhaolong Wu[1,2], Shuwen Zhang[1,2], Wei Li Wang[1,3,4], Yinping Ma[1], Yuanchen Dong[5] & Youdong Mao[1-4]*

[1]State Key Laboratory for Mesoscopic Physics, School of Physics, Peking University, Beijing, China. [2]Center for Quantitative Biology, Peking University, Beijing, China. [3]Intel Parallel Computing Center for Structural Biology, Dana-Farber Cancer Institute, Boston, MA, USA. [4]Department of Cancer Immunology and Virology, Dana-Farber Cancer Institute, Boston, MA, USA. [5]Institute of Chemistry, Chinese Academy of Sciences, Beijing, China.

*To whom correspondence may be addressed.

## Abstract

The 2.5-MDa 26S proteasome maintains proteostasis and regulates myriad cellular processes[1]. How polyubiquitylated substrate interactions regulate proteasome activity is not understood[1,2]. Here we introduce a deep manifold learning framework, named AlphaCryo4D, which enables atomic-level cryogenic electron microscopy (cryo-EM) reconstructions of nonequilibrium conformational continuum and reconstitutes 'hidden' dynamics of proteasome autoregulation in the act of substrate degradation. AlphaCryo4D integrates 3D deep residual learning[3] with manifold embedding[4] of energy landscapes[5], which directs 3D clustering via an energy-based particle-voting algorithm. In blind assessments using simulated heterogeneous cryo-EM datasets, AlphaCryo4D achieved 3D classification accuracy three times that of conventional methods[6-8] and reconstructed continuous conformational changes of a 130-kDa protein at sub-3-Å resolution. By using AlphaCryo4D to analyze a single experimental cryo-EM dataset[2], we identified 64 conformers of the substrate-bound human 26S proteasome, revealing conformational entanglement of two regulatory particles in the doubly capped holoenzymes and their energetic differences with singly capped ones. Novel ubiquitin-binding sites are discovered on the RPN2, RPN10 and α5 subunits, which remodel polyubiquitin chains for deubiquitylation and recycling. Importantly, AlphaCryo4D choreographs single-nucleotide-exchange dynamics of proteasomal AAA-ATPase motor during translocation initiation, which upregulates proteolytic activity by allosterically promoting nucleophilic attack. Our systemic analysis illuminates a grand hierarchical allostery for proteasome autoregulation.

The 26S proteasome is the known largest ATP-dependent protease machinery of ~2.5 MDa molecular weight ubiquitously found in all eukaryotic cells. It exists at the center of the ubiquitin-proteasome system (UPS) that regulates myriad cellular processes, such as protein quality control, cell cycle, gene expression, inflammation and immunity[1,2]. Dysregulation of the UPS has been broadly associated with carcinogenesis, ageing, neurodegenerative and cardiovascular diseases[1]. The 26S holoenzyme is assembled by a cylindrical 20S core particle (CP) capped with one or two 19S regulatory particles (RPs), each consisting of the lid and base subcomplexes[1,2]. The ring-like heterohexameric motor of ATPases-associated-with-diverse-cellular-activities (AAA) ATPase in the base subcomplex mechanically unfold polyubiquitylated substrates recognized by ubiquitin receptors in the lid subcomplex. Previous structural data by cryogenic electron microscopy (cryo-EM) have suggested three modes of coordinated ATP hydrolysis in the AAA-ATPase motor that regulate distinct steps of ubiquitin recognition, deubiquitylation, initiation of translocation and processive substrate translocation in the proteasome[2]. However, key intermediate conformations between these states were missing, greatly limiting our understanding of the intrinsic proteasome regulation. The mechanism of the proteasome targeting of substrate-conjugated ubiquitin signals is not fully understood[1,9-11]. How polyubiquitylated substrate interactions regulate proteasome activity remains enigmatic[1].

The 26S proteasome is one of the most complex, dynamic and conformationally heterogeneous holoenzyme machinery in cells[1,2,12-15], challenging the current approaches in cryo-EM structure determination[6-9,16-22]. Visualizing atomic structures of transient, nonequilibrium intermediates connecting the major states of the proteasome has been unsuccessful by conventional cryo-EM analysis[1,2,7,15,17,18]. To address this major challenge, we developed a novel deep manifold learning system codenamed AlphaCryo4D that can break such a limitation and enable atomic-level cryo-EM reconstructions of highly dynamic, lowly populated intermediates or transient states. By applying AlphaCryo4D to analyze a large cryo-EM dataset that previously solved the atomic structures of 7 proteasomal RP-CP states[2], we reconstructed 64 conformers of the human 26S proteasome in the act of substrate degradation up to 2.5-Å resolution. Systematic analyses of those transient or intermediate states in substrate processing pinpoint energetic differences between the doubly and singly capped proteasomes, discover several novel ubiquitin-binding sites in the proteasome, and choreograph single-nucleotide-exchange dynamics of the proteasomal AAA-ATPase motor during translocation initiation. Importantly, the dramatically enriched proteasomal states provide unprecedented insights into an integrative, hierarchical allosteric mechanism of proteasome autoregulation upon polyubiquitylated substrate interactions.

**Design of deep manifold learning**

The conceptual framework of AlphaCryo4D integrates unsupervised deep learning with manifold embedding to learn a free-energy landscape, which directs cryo-EM reconstructions of conformational continuum or transient states via an energy-based particle-voting algorithm. AlphaCryo4D involves four major steps (Fig. 1a, Extended Data Fig. 1, Methods): (1) Hundreds to thousands of 3D volumes are bootstrapped with $M$-fold particle shuffling and Bayesian clustering in RELION[6,7,17,18], during which each particle is used $M$ times in reconstructing $M$ volumes; each of the $M$ copies is called a 'vote' of the particle (Extended Data Fig. 1a). Due to $M \geq 3$, this step demands a higher computational cost as compared with conventional methods[6-8,17,18]. (2) 3D feature maps of all volumes are learnt by a 3D deep residual autoencoder[3,23] and are mapped to a manifold by t-distributed stochastic neighbor embedding[4,5], which is used to compute an energy landscape through the Boltzmann relation (Fig. 1b, d, e, Extended Data Table 1). (3) A string method[24] is used to search the minimum-energy path (MEP) on the energy landscape[25], which defines local energy minima or transition states as the centers of clustering boundaries for particle voting (Fig. 1f). (4) Each particle is counted for the number of votes that are casted within the voting boundaries on the energy landscape. The particle is classified to the cluster that receives more than $M/2$ votes of this particle within its voting boundary (Fig. 1f, Extended Data Fig. 1b). Particles that cannot be reproducibly voted into any clustering boundaries for more than $M/2$ times are 'voted out'. The resulting 3D classes are expected to be conformationally homogeneous enough for high-resolution cryo-EM refinement in RELION[6] or cryoSPARC[8]. Since many protein complexes exhibit profound conformational changes in different local regions, we also implemented a focused classification strategy of AlphaCryo4D that applies a local 3D mask[17] throughout the entire procedure, often being executed as an iterative step after initial 3D classification by AlphaCryo4D in the absence of any 3D mask (Fig. 1a, Methods).

**Solving conformational continuum at atomic level**

To assess the numerical performance of AlphaCryo4D, we generated three large synthetic heterogeneous cryo-EM datasets with signal-to-noise ratios (SNRs) of 0.05, 0.01 and 0.005. Each dataset includes 2 million randomly oriented single particles computationally simulated from 20 hypothetical conformer models of the ~130-kDa NLRP3 inflammasome protein[26]. These conformers imitate conformational continuum of the NACHT domain rotating around the LRR domain over an angular range of 90° during inflammasome activation[26] (Fig. 1c). We conducted blind assessments on 3D classification and heterogeneous reconstructions by AlphaCryo4D, without providing any information of particle orientations, translations and conformational identities (Fig. 1d-f, Extended Data Fig. 2, Methods). The 3D classification precision of a retrieved conformer was computed as the ratio of the particle number of correct class assignment (based on the ground truth) versus the total particle number in the class. The results were then compared with several alternative methods in 18 blind tests, including conventional maximum-likelihood-based 3D (ML3D) classification in RELION[6,7,17], 3D variability analysis (3DVA) in cryoSPARC[8,21], and deep generative model-based cryoDRGN[22]. In all blind tests, AlphaCryo4D retrieved all 20 conformers and markedly outperformed the alternative methods, with an average of 3D classification precision at 0.83, 0.82 and 0.65 for datasets with SNRs of 0.05, 0.01 and 0.005, respectively (Fig. 2a, Extended Data Fig. 3a). By contrast, all alternative methods missed two to seven conformers entirely and exhibited 3D classification precisions in the range of 0.2-0.5 in general (Fig. 2b-d, Extended Data Figs. 3b-j).

The 3D classification precision appears to be strongly correlated with the map quality and resolution homogeneity across the density map (Fig. 2, Extended Data Fig. 2h). All density maps from AlphaCryo4D consistently show homogeneous local resolutions (at 2.6-2.9 Å for SNR of 0.01) between the NACHT and LRR domains (Fig. 2e, i, Extended Data Figs. 2h, 3k, 4a). By contrast, all density maps by the alternative methods show lower average resolutions and notably heterogeneous local resolutions, with the NACHT domain exhibiting substantially lower resolution than that of the LRR domain, causing blurred features, broken loops and invisible sidechains in NACHT (Fig. 2f-h, j-l, Extended Data Figs. 3l-n, 4b-d). Thus, the significantly improved 3D classification accuracy by AlphaCryo4D enables 4D reconstructions of conformational continuum at the atomic level.

To understand how the 3D classification accuracy is improved, we further studied the algorithmic mechanism by removing or replacing certain components of AlphaCryo4D in 24 control tests (Extended Data Figs. 2d-g, 5, see Methods for details). By removing the component of deep manifold learning from AlphaCryo4D, the statistical distributions of high-precision 3D classes after particle voting are reduced by ~60% (Extended Data Fig. 5d-f). Similarly, classifying particles without particle voting or in the absence of deep residual autoencoder reduces the high-precision 3D classes by ~15-30% at a lower SNR (Extended Data Fig. 5m-r). Throughout AlphaCryo4D, the steps of particle shuffling, defining voting boundaries on energy landscapes and energy-based particle voting contribute to ~16%, ~20% and ~40% improvements of high-precision 3D classes, respectively (Extended Data Fig. 5g-l, s-u). Taken together, these results indicate that the performance of deep manifold learning is synergistically enhanced by energy-based particle voting.

**Visualizing hidden dynamics of the proteasome**

Having conducted the proof-of-principle study of AlphaCryo4D using the simulated datasets, we then turned to several experimental datasets[2,14]. We first applied this approach to analyze an existing cryo-EM dataset of the substrate-free ATPγS-bound 26S proteasome[14]. The AlphaCryo4D-reconstructed energy landscape clearly shows six local energy minima corresponding to the known six conformational states $S_A$, $S_B$, $S_C$, $S_{D1}$, $S_{D2}$ and $S_{D3}$, verifying the applicability of AlphaCryo4D in processing real experimental data[14] (Fig. 3a). Next, we used AlphaCryo4D and its focused classification procedure to analyze a larger cryo-EM dataset of the substrate-engaged 26S proteasome (Fig. 3b, c, Extended Data Fig. 6 and Methods)[2]. This dataset was collected on the human 26S proteasome mixed with a Lys63-linked polyubiquitinated substrate Sic1[PY] for 30 seconds before ATP was diluted with ATPγS, which was expected to maximally capture any intermediate states before the degradation reaction was completed[2]. The reconstruction of the energy landscape of the substrate-engaged proteasome suggests that the rate-limiting step of substrate degradation lies in the initiation of substrate unfolding after substrate deubiquitylation, reflected in the highest activation energy barrier between states $E_B$ and $E_{C1}$ (Fig. 3b, c), consistent with a recent fluorescence microscopy study on the proteasome kinetics[27].

Impressively, AlphaCryo4D extracted significantly more conformers of the proteasomes in various forms from the same cryo-EM dataset that previously yielded seven atomic structures of the proteasome[2]. It discovered 20 conformational states of RP-CP subcomplex (Fig. 3b, c,

Extended Data Figs. 6, 7, Extended Data Table 2). Nine transient states (designated $E_{A1.1}$, $E_{A1.2}$, $E_{A2.2}$, $E_{A2.3}$, $E_{B.2}$, $E_{B.3}$, $E_{D2.2}$, $E_{D2.3}$ and $E_{D.\alpha5}$) at 3.1-7.5 Å resolution exhibit previously unseen features resembling ubiquitin that map various putative ubiquitin-binding sites on the RPN1, RPN2, RPN10 and α5 subunits in the proteasome (Fig. 4, Extended Data Fig. 6p, u). Six sequential intermediate states (designated $E_{D0.1}$, $E_{D0.2}$, $E_{D0.3}$, $E_{D1.1}$, $E_{D1.2}$ and $E_{D1.3}$) at 3.2-3.9 Å resolution bridge a major missing gap between states $E_{C2}$ and $E_{D1}$, revealing how concerted structural transitions in the AAA-ATPase motor during single-nucleotide exchange drive the initiation of substrate unfolding (Fig. 5).

To understand whether the singly capped (SC) and doubly capped (DC) proteasomes behave differently, AlphaCryo4D reconstructed 8 and 36 conformational states for the pure SC and DC holoenzymes, respectively (Fig. 3d, e, Extended Data Figs. 8, 9a, b). All 8 SC and 29 DC states were refined to 3.7-4.7 Å, with 7 DC states limited to 6-9 Å due to their transient nature and extremely low population (Extended Data Fig. 8b, e, f). Note that the CP gate in the RP-distal side of the 20 states of RP-CP subcomplex is in a heterogeneous conformation of various stoichiometric ratios between closed and open states. By contrast, both CP gates in the 44 states of pure SC or DC reconstructions are in a homogeneous conformation.

Besides resolving significantly more conformers, AlphaCryo4D pushes the envelope of the achievable resolution of known states due to its advantage in keeping more particles without sacrificing conformational homogeneity. For example, the structure of state $E_{D2}$ was improved from previous resolution at 3.2 Å to 2.5 Å, allowing us to investigate how the CP gating allosterically regulates the catalytic activity of the β-type subunits (Fig. 6, Extended Data Fig. 9c-h). The local resolution of RPN11-bound ubiquitin in state $E_{A2}$ was also improved from 5-6 Å to 3.5 Å among many other improvements (Extended Data Fig. 7a).

**Conformational entanglement**

Both the SC and DC proteasomes are abundant in cells[28]. Their structural and functional difference, however, remains elusive. In the 36 DC reconstructions, each conformer is a combination of two RP states. Each CP gate in the DC proteasome appears to be rigorously controlled by their respective proximal RP and is only open when its proximal RP is in an $E_D$-compatible state with five RPT C-terminal tails inserted into the surface pockets on the α-ring[2] (Fig. 3e, Extended Data Fig. 8a-d). Consistently, the gate on the RP-free side of the CP in the SC proteasome is closed in all states, indicating that the CP gate opening on the RP-controlled side does not remotely open the other CP gate on the RP-free side (Extended Data Fig. 8d).

To understand whether the conformational states of two RPs in the DC proteasome allosterically influence each other, we compare the state distribution matrix obtained by AlphaCryo4D to the prediction of a control model assuming two completely uncoupled RP states in the DC proteasome (Fig. 3f-h). Although the overall pattern of the experimental distribution roughly agrees with the model prediction, a notable number of DC states exhibit substantial deviation, indicating that the conformational dynamics of the two RPs in the same proteasome are profoundly entangled together (Fig. 3g, h). Intriguingly, the most prominent positive deviations are seen in the state distributions of $E_{A1}$-$E_{A1}$, $E_B$-$E_B$, $E_B$-$E_{D2}$, and $E_{D2}$-$E_{D2}$ combinations (Fig. 3g), suggesting that the deubiquitylation state $E_B$ and the translocation state $E_{D2}$ of one RP are allosterically coupled to and mutually upregulated by states $E_B$ and $E_{D2}$ of the opposite RP. Coupled actions of deubiquitylation and translocation between two opposite RPs are expected to optimize the efficiency of substrate processing. Moreover, all the DC states along the diagonal of the state distribution matrix are more or less upregulated (indicated by red in Fig. 3g, h), suggesting that both RPs in the same proteasome synergistically promote each other to transit into downstream states in a symmetrical fashion.

Notably, the largest discrepancy of the distribution of RP states between the SC and DC proteasomes also lies in state $E_B$ (Fig. 3b-d). The state population of $E_B$ is significantly lower in the SC than in the DC proteasome. This observation verifies the conclusion drawn from the above analysis on the DC proteasomes alone that state $E_B$ is upregulated by the conformational entanglement effect in the DC holoenzyme, suggesting that the SC proteasome lacks the functional synergy between two opposite RPs seen in the DC proteasome (Fig. 3h). Altogether, these results implicate that the DC proteasome could be energetically more efficient in substrate deubiquitylation and degradation than the SC proteasome.

**Ubiquitin-proteasome interactions**

Unexpectedly, in states $E_{A1.2}$, $E_{A2.2}$, and $E_{B.2}$, a diubiquitin-like density is consistently found on RPN2, whereas a monoubiquitin-like density is observed at the apex of the von Willebrand factor type A (VWA) domain[29] of RPN10 in states $E_{A2.3}$, $E_{B.3}$ and $E_{D2.3}$ (Fig. 4a, Extended Data Fig. 7b-f). In state $E_{D2.2}$, the diubiquitin-like density, however, appears to contact both RPN2 and RPN10, and can be modelled with a Lys63-linked diubiquitin (Extended Data Fig. 7b, g). The RPN10 ubiquitin-interacting motifs (UIMs), however, were not reliably observed presumably due to their highly flexible nature (Extended Data Fig. 7g). The populations of these states are very small and represented in only a few thousand particles or 0.1-0.3% of the entire dataset, indicating their transient nature. The local resolutions of the RPN2- and RPN10-bound ubiquitin-like densities are lower than the rest of the reconstructions and other ubiquitin densities on RPN1, RPN11 and α5 subunits. Rigid-body fitting of the ubiquitin structure into these ubiquitin-like densities confirms the agreement of the overall density shape with the ubiquitin fold (Fig. 4a, Extended Data Fig. 7b-g).

**RPN1 sites**. By focused 3D classification via AlphaCryo4D, we refined the density map of RPN1 in state $E_{A1.0}$ (~5.9% of the dataset) to 3.7 Å, which was previously reconstructed at moderate resolution (~5.5 Å)[2] (Extended Data Figs. 6n, 7d). It reveals that a putative ubiquitin moiety contacts Glu458, Cys459 and Asp460 at the N-terminal end of an RPN1 helix (homologous to H24 between the T1 and T2 sites of yeast Rpn1) and potentially interacts with the T2 site (Asp423, Leu426 and Asp430) on the RPN1 toroidal domain (Fig. 4c, d)[11]. Because the yeast Rpn1 T2 site was found to preferentially bind Ubp6 ubiquitin-like (UBL) domain[11], this site may be sterically occluded for ubiquitin binding in the presence of USP14/Ubp6. In addition, AlphaCryo4D also identified another state $E_{A1.1}$ that exhibits a diubiquitin-like density with one ubiquitin moiety on the RPN1 T1 site and the other near the T2 site (Extended Data Fig. 7e). The much smaller population of state $E_{A1.1}$ (~1.3 % of the dataset) is consistent with the previous finding that Lys63-linked polyubiquitin chains have a lower affinity against the T1 site than those linked via Lys48 and Lys6 residues[11].

**RPN2 site**. The ubiquitin-binding site on RPN2 is centered on three helices harboring an acidic surface around residues Tyr502, Glu530, Asp531 and Ser569 (Fig. 4b-e). Structural comparison suggests that the ubiquitin-binding site on RPN2 is homologous to the T1 site of RPN1 (Fig. 4c, d)[11]. We therefore name it 'the RPN2 T1 site'. Although there is little conservation in amino acid sequences between the RPN1 and RPN2 T1 sites, their structural and electrostatic properties are highly conserved (Fig. 4d, Extended Data Fig. 7h-k). Comparison with the NMR structure of Lys48-linked diubiquitin-bound RPN1 T1 site helices[11] further confirms the conservation and similarity of the ubiquitin-binding modes between the RPN1 and RPN2 T1 sites (Fig. 4c). Consistent with this finding, previous studies have observed weak interactions of UBL proteins and RPN2 by cross-linking experiments[30]. As there is also ubiquitin bound to RPN11 in states $E_{A2.2}$, and $E_{B.2}$, a Lys63-linked triubiquitin chain can be fitted into the RPN2- and RPN11-bound ubiquitin densities by modeling the rotation of Lys63 linkages, suggesting that the RPN2 T1 site may function as a coreceptor site for ubiquitin transfer from RPN10 to RPN11 to facilitate deubiquitylation (Fig. 4g). Given that at least one RPN10 UIM-bound ubiquitin invisible in these cryo-EM maps would be expected to exist, the observation of triubiquitin on RPN2 and RPN11 structurally rationalizes why tetraubiquitin is an optimal degradation signal for the proteasome[1].

**RPN10 VWA site**. In states $E_{A2.3}$, $E_{B.3}$ and $E_{D2.3}$, the monoubiquitin-like densities consistently reside on Lys132, Lys133 and Lys163 of the RPN10 VWA domain. By contrast, in state $E_{D2.2}$, a diubiquitin-like density appears to span from this lysine-rich site of RPN10 to the RPN2 T1 site and can be modelled with a Lys63-linked diubiquitin (Extended Data Fig. 7b, g). In line with these observations, previous studies have suggested that the yeast Rpn10 VWA is a ubiquitin-binding domain and the ubiquitylation of Rpn10 VWA itself prevents Rpn10 from being incorporated into the proteasome[29]. As ubiquitylation of RPN10 was found to regulate substrate recruitment[29,31-33], it remains to be clarified whether the putative ubiquitin at this lysine-rich site is covalently modified by ubiquitylation of RPN10 itself co-purified from of HEK293 cells.

**α5-subunit site**. In state $E_{D2.2}$, we can observe a ubiquitin density bound to the RPN1 T2 site and another weaker ubiquitin density at the α5-subunit ~6 nm beneath the RPN1 T2 site (Fig. 4a, Extended Data Fig. 7b). To understand if the ubiquitin-binding site on the α5 subunit is used more broadly, we combined all the $E_D$-compatible states and focused the 3D classification of AlphaCryo4D on the α5-bound ubiquitin. This yielded a much larger 3D class, named state $E_{D.α5}$ (~4% of the entire dataset) that improved the ubiquitin-bound CP structure to 3.1 Å, and the α5-bound ubiquitin density to 4.9 Å (Extended Data Figs. 6o, p, r, u, 7f). The ubiquitin-binding site on the α5 subunit is composed of acidic or hydrophobic residues Glu183, Val184, His186, Ser188 and Glu193 in a loop connecting two helices, highly resembling the mode of ubiquitin recognition by an RPN13 loop[10,34] (Fig. 4f, Extended Data Fig. 7f). The electrostatic property of this site is also similar to those of the RPN1 T1/T2 and RPN2 T1 sites that are all acidic, in charge complementarity with the basic ubiquitin surface around residues Ile44 and His68 (Extended Data Fig. 7h-o). Although no previous evidence supports ubiquitin interactions with the CP, recent data suggest that the free 20S CP can degrade substrate-conjugated ubiquitin, which might be facilitated by ubiquitin binding to the α-ring[35].

We note that the putative ubiquitin-binding sites reside in the vicinity of primary receptor sites. The RPN2 T1 site is very close to the RPN10 UIMs, whereas the α5 site is near the RPN1 T2 site. Modelling of a Lys63-linked tetraubiquitin chain by fitting its terminal ubiquitin into the cryo-EM density suggests that a single tetraubiquitin can span across the α5 subunit and the RPN1 T2 site (Fig. 4h). Given that RPN1 appears to allosterically regulate the AAA-ATPase motor conformation[2] (Fig. 5a, b), tetraubiquitin binding to both the RPN1 T2 site and α5 subunit could allosterically stabilize the open-CP states ($E_{D0}$, $E_{D1}$ and $E_{D2}$). It is unclear whether these low-affinity ubiquitin-binding sites would directly involve in primary targeting of substrate-conjugated polyubiquitin chains by the proteasome in cells. However, it is conceivable that the polyubiquitin chains could be first captured by the high-affinity sites and then moved to the nearby low-affinity sites. Working together, these ubiquitin-binding sites could remodel polyubiquitin chains and deliver the peptide-proximal ubiquitin to the deubiquitylation site[2] (Fig. 4g) as well as assist ubiquitin recycling after isopeptide bond hydrolysis (Fig. 4h). Due to the limited ubiquitin resolution and transient nature of these ubiquitin-bound states, we cannot rule out that some of the ubiquitin-like densities are co-purified endogenous ubiquitin, UBL proteins or ubiquitylation of the proteasome itself[1,36]. Future functional studies are required to understand their definitive chemical nature and biological importance in cells.

**Single-nucleotide exchange dynamics**

The six intermediate states ($E_{D0.1}$, $E_{D0.2}$, $E_{D0.3}$, $E_{D1.1}$, $E_{D1.2}$ and $E_{D1.3}$) during initiation of substrate unfolding and translocation exhibit highly coordinated movements in the RP relative to the CP (Fig. 5a-c, Extended Data Fig. 10), with prominent rotations in the RPN1 N-terminal that detaches from RPT2 in states $E_{D0.1}$, $E_{D0.2}$, and $E_{D0.3}$ and re-binds RPT2 in state $E_{D1.1}$, $E_{D1.2}$ and $E_{D1.3}$. All six conformations share the same architecture in their substrate interactions with pore-1 loops from four RPT subunits (RPT2, RPT6, RPT3 and RPT4), which remain largely unchanged in overall conformation (Fig. 5d). These structures illustrate that the RPT2 re-association with the substrate proceeds that of RPT1 after both RPT1 and RPT2 are dissociated from the substrate in state $E_C$. They also share the same pattern of nucleotide states, with ATP bound to RPT1, RPT2, RPT6 and RPT3 and ADP bound to RPT4. There are poor, partial nucleotide densities in the RPT5 nucleotide-binding pocket, which give rise to the highest B-factor (250-500) when fitted with ADP, suggesting that RPT5 undergoes ADP release in these states (Extended Data Fig. 10k).

Markedly, the pore-1 loop of RPT1 gradually moves toward the substrate from a distance of ~16 Å to ~3 Å, whereas the pore-1 loop of RPT5 gradually moves away from the substrate from a distance of ~3 Å to ~16 Å (Fig. 5d, f). The small AAA subdomain of RPT5 moves inward together with the large AAA subdomain of RPT1, as the large AAA subdomain of RPT5 rotates away from the substate (Extended Data Fig. 10f-i). Consistently, the arginine fingers (R-finger) of RPT2 are progressively moved toward ATP bound in RPT1, closing up the nucleotide-binding pocket (Fig. 5e, g). In contrast, the nucleotide-binding pocket of RPT5 is gradually shrunk but is not fully closed by the R-fingers of RPT1, whereas the nucleotide-binding pocket of RPT4 is gradually opened up (Fig. 5g). As a result, the pore-1 loops of RPT1 and RPT5 are only in contact with the substate in the last and first states $E_{D1.3}$ and $E_{D0.1}$, respectively. Both are dissociated from the substate in states $E_{D0.2}$, $E_{D0.3}$, $E_{D1.1}$, and $E_{D1.2}$. These gradual movements allow us to assign their temporal sequences as illustrated (Fig. 5a-e), indicating that these states represent intermediate conformations accompanying single-nucleotide exchange that occurs in RPT5 (Fig. 5h). Interestingly, the outward flipping of RPT5 appears to complete its major conformational change in state $E_{D0.3}$, much faster than the inward motion of RPT1, which completes its conformational change in state $E_{D1.3}$. This observation suggests that ATPase re-engagement with the substrate is the rate-limiting step in the single-nucleotide exchange kinetics of the AAA-ATPase motor.

Our data clarify that the CP gate is open once the pore-1 loop of RPT2 binds the substrate in state $E_{D0.1}$, as five RPT C-tails are inserted in the α-pockets in all six states (Extended Data Fig. 8c). Altogether, these structures vividly characterize at the atomic level how the "hand-over-hand" actions of RPT subunits are executed by coordinated conformational dynamics in the AAA-ATPase motor during single-nucleotide exchange and generates mechanical force unfolding the substrate. Although the sequential model of coordinated ATP hydrolysis around the AAA-ATPase ring has been hypothesized in numerous similar systems[1,2,12,15], our results provide the first direct evidence for sequential intermediate states of hand-over-hand actions within a single cycle of nucleotide exchange.

**Hierarchical allosteric regulation of proteolytic activity**

Both atomic structures of the substrate-bound CPs in the open- and closed-gate states $E_A$ and $E_{D2}$ show a segment of substrate polypeptide binding the catalytically active residue Thr1 in the β2 subunit, whereas no substrate density is observed in the other two catalytic subunits β1 and β5. In both states, the substrate is bound in a β-sheet conformation between residues 20 to 22 and 46 to 48 in the β2 subunit (Fig. 6a, b, Extended Data Fig. 9e, f), reminiscent of the conformations of CP inhibitors binding the β5 subunits in the crystal structures of the CP[37,38]. The hydroxyl oxygen atom in the catalytic Thr1 is ~2-Å closer to the substrate mainchain in state $E_{D2}$ compared to state $E_A$. This structural difference is expected to result from the CP gate opening that exerts a profound allosteric impact on the entire CP structure[39-42].

To understand how such a subtle structural change influences the proteolytic activity, we carried out first-principles calculations of the quantum-mechanical interactions involved in the nucleophilic attack of the catalytic residue Thr1 on the scissile peptide bond in the substrate, using the density functional theory (DFT) with the generalized gradient approximation[43]. To initiate the nucleophilic attack, the carbonyl group needs to be brought to the proximity of the electrophile Thr1-Oγ to facilitate formation of a tetrahedral transition state[38,44,45]. We calculated the charge density difference induced around the interacting pair after proton transfer from Thr1-Oγ to Thr1-$NH_2$ (Fig. 6c, d, Extended Data Fig. 9g, h). The substrate-interacting pair of atoms in state $E_{D2}$ form a bonding region of slight positive charge difference linking the carbonyl group of the substrate to Thr1-Oγ. By contrast, the pair are separated by a charge depletion non-bonding region in state $E_A$. Indeed, a recent study has found that the nucleophilic attack is the rate-limiting step in the acylation barrier[44]. Our result suggests that CP gate opening allosterically promotes the nucleophilic attack by lowering the activation energy barrier of acylation and induces hybridization of molecular orbitals, hence upregulating the reaction rate of proteolysis. In support of our structural finding, previous studies have observed indirect evidence for coupling of the CP gating with the proteolytic activity[39-42].

Taken together, our systematic structural analyses point to a grand allostery model for the proteasomal autoregulation of proteolytic activity through a three-tier hierarchy of inter-molecular interaction networks (Fig. 6e), in which all subunit conformations are allosterically coupled. The first tier of autoregulation is between the CP gate in the α-ring and the catalytic sites of the β-subunits. Opening of the CP gate allosterically upregulates the proteolytic activity of the three β-subunits. The second tier of autoregulation lies between the substrate-bound AAA-ATPase motor and the rest of the proteasome. The AAA-ATPase motor bridges the ubiquitin-binding sites and the CP via direct interactions of the RPT coiled-coil (CC) domains with RPN1, RPN2 and RPN10 harboring ubiquitin-binding sites. As a result, three modes of coordinated ATP hydrolysis appear to regulate the intermediate steps of substrate processing[1,2], in which initiation of substrate unfolding is allosterically coupled with CP gating during $E_{C2}$ to $E_{D0.1}$ transition (Fig. 5). The third tier of autoregulation is via dynamic interactions between ubiquitin signals and a collection of ubiquitin receptor sites and deubiquitinases. This is manifested as the dual roles of ubiquitin-binding sites in recognizing ubiquitin signals and in transmitting such recognitions to allosteric regulation of AAA-ATPase conformations that control the CP gate. Ubiquitin binding and remodeling destabilizes the resting state of the proteasome and promotes its transition to states $E_B$, $E_C$ and ultimately $E_D$. On the other hand, a polyubiquitin chain binding multiple sites in the proteasome could have profound allosteric and conformational selection effects (Fig. 4g, h). Such three-tier, hierarchical allosteric regulations may occur mutually with both RPs and are expected to underlie the conformational entanglement of the two RPs in the DC proteasome (Fig. 3e), which was not clearly observed in the absence of substrates[1,13,14]. This grand allostery mechanism might have been evolved to optimize the proteasome function in response to extremely complicated intracellular challenges and is manifested as various phenotypic effects, such as UBL-induced proteasome activation[46,47] and long-range allosteric couplings in the proteasome[1,40,41,48,49].

**Concluding remarks**

In this work, motivated by solving the hidden dynamics of proteasome autoregulation, we developed AlphaCryo4D, a novel framework of deep manifold learning, that enables 3D reconstructions of nonequilibrium conformational continuum at the atomic level. In retrospect, mapping transient ubiquitin-proteasome interactions, choreographing sequential intermediates within a single cycle of nucleotide exchange, and detecting conformational entanglement of the two RPs associated with the same CP as well as subtle allosteric effects of polypeptide hydrolysis by CP gating, each itself presents a formidable challenge. Being able to accomplish all these missions by solving 64 proteasomal conformers from a single experimental dataset suggests that AlphaCryo4D is a significant advance toward solving 3D heterogeneity problems in cryo-EM structure analysis at the atomic level. More importantly, AlphaCryo4D has revealed 'previously invisible' mechanistic details in nonequilibrium dynamics of the proteasome autoregulation in the act of polyubiquitylated substrate degradation and offers unprecedented insights into the grand hierarchical allostery underlying proteasome function. With recent improvements in cryo-EM instruments[19,20] and time-resolved sample preparation methods[50], we expect that further development and widespread applications of AlphaCryo4D or similar approaches will boost cryo-EM imaging as a de novo discovery tool and transform modern biomedical research.

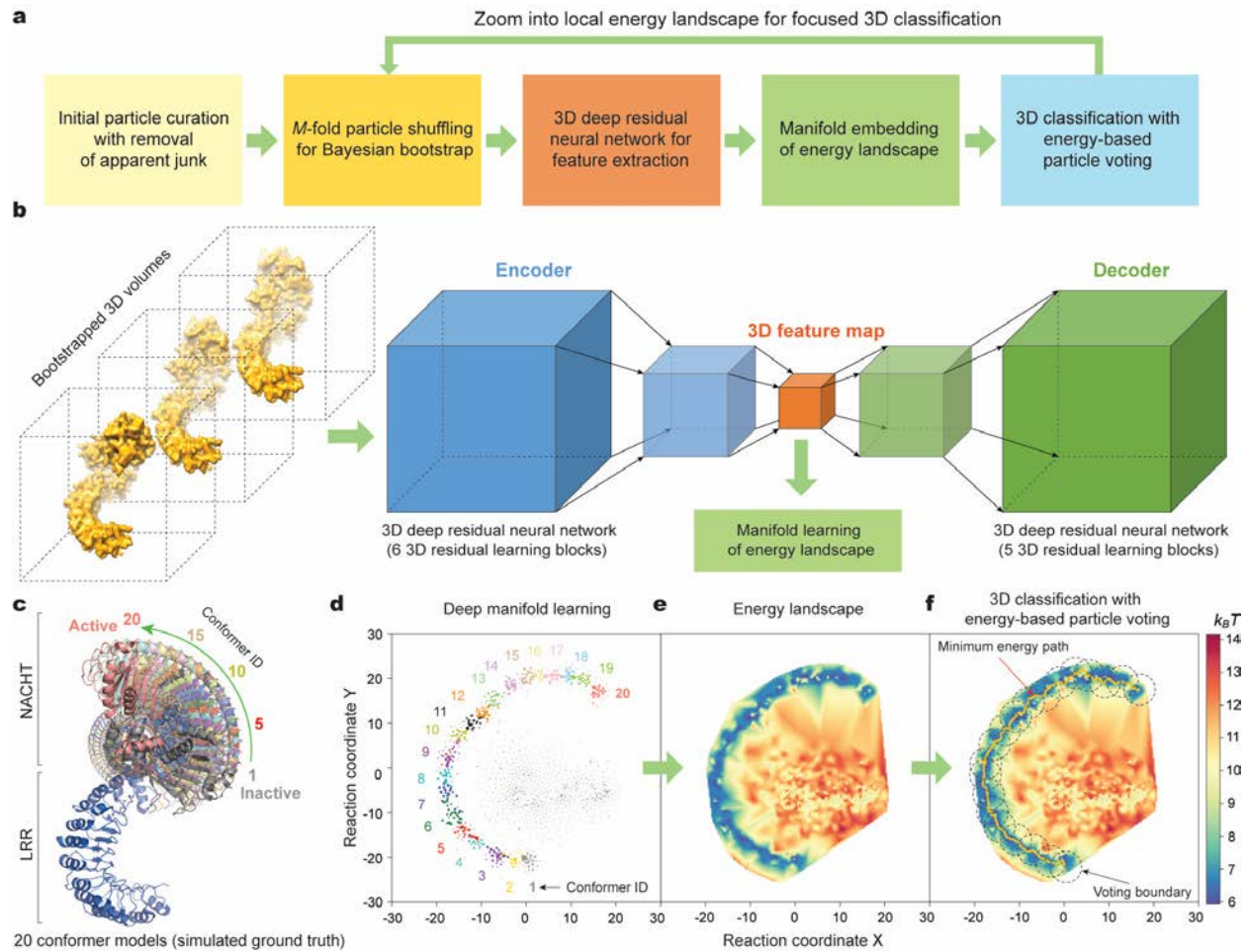

**Figure 1. Conceptual framework of AlphaCryo4D for 4D cryo-EM reconstruction. a**, Schematic showing the major conceptual steps of single-particle cryo-EM data processing in AlphaCryo4D. **b**, Illustration of deep residual learning of 3D feature maps by an autoencoder conjugated to a decoder for unsupervised training. **c**, The 20 atomic models of hypothetical conformers of NLRP3 in cartoon representations simulate a 90° rotation of the NACHT domain relative to the LRR domain based on the structure of NLRP3 (PDB ID: 6NPY). **d**, Manifold learning of the bootstrapped 3D volumes and their feature maps learnt by the autoencoder. Each data point corresponds to a 3D volume. The color labels the conformer identity of ground truth for the purpose of verification. **e**, Energy landscape computed from the manifold shown in **d** using the Boltzmann distribution. **f**, Minimum energy path (orange line) calculated by the string method is used to find the approximate cluster centers of 20 conformers. The cluster boundaries for energy-based particle voting are shown as dashed circles.

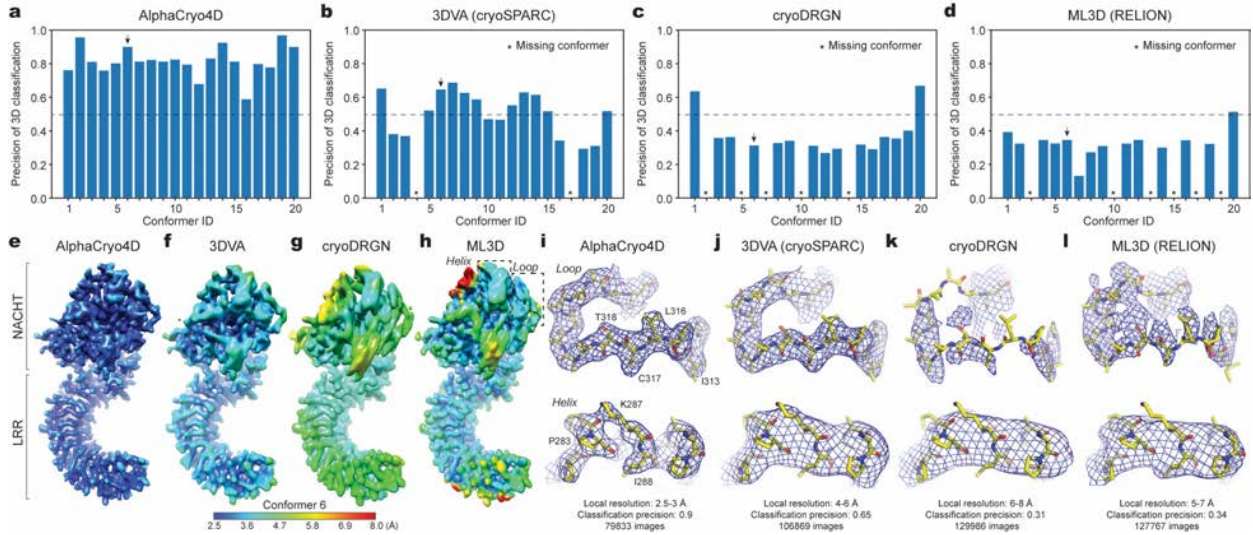

**Figure 2. Performance evaluation of AlphaCryo4D for reconstructions of conformational continuum at the atomic level**. **a-c**, Plots of 3D classification precisions of the 20 NLRP3 conformers from blind assessments on the simulated NLRP3 dataset with SNR of 0.01, using AlphaCyo4D (**a**), 3DVA algorithm in cryoSPARC[21] (**b**), cryoDRGN[22] (**c**), and conventional ML3D in RELION[6] (**d**). All 3D conformers in panel (a) were reconstructed to 2.6-2.9 Å resolution (Extended Data Fig. 3m). Asterisks mark the missing conformers that were completely lost due to misclassification by 3DVA, cryoDRGN and ML3D. **e-h**, Typical side-by-side comparison of density map quality and local resolution of the same conformer (ID 6) reconstructed by AlphaCryo4D (**e**), 3DVA (**f**), cryoDRGN (**g**) and ML3D (**h**). The maps are colored by their local resolutions calculated by Bsoft blocres program. **i-l**, Closeup side-by-side comparison of the same two secondary structures, including a loop (upper row) and a helix (lower row), in the NACHT domain illustrates considerable improvements in local density quality and resolution by AlphaCryo4D (**i**) as opposed to 3DVA/cryoSPARC (**j**), cryoDRGN (**k**) and ML3D/RELION (**l**). The locations of the loop and helix in the NLRP3 structure are marked by dashed boxes in panel (**h**). The same ground-truth atomic model of Conformer 6 shown in stick representations is superimposed with the density maps shown in blue mesh representations, all from the same perspective. The atomic model was not further refined against each map for visual validation of the map accuracy.

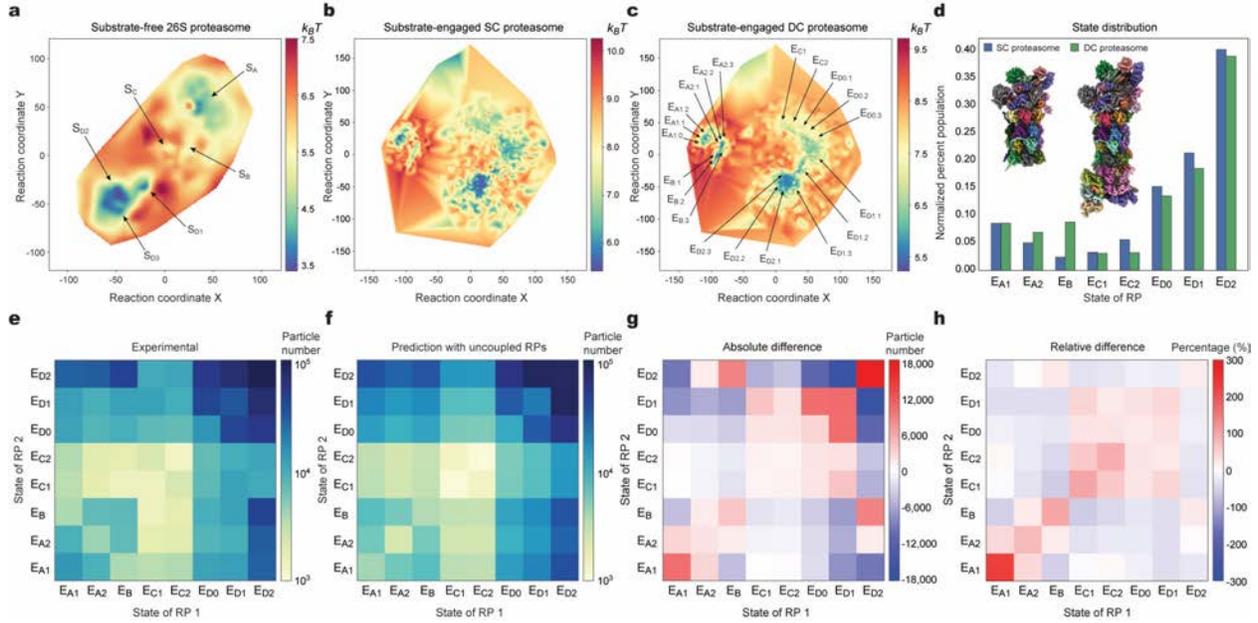

**Figure 3**. **Energetic differences of the singly and doubly capped human proteasomes**. **a-c**, Energy landscape of the substrate-free ATPγS-bound 26S proteasome (**a**), the substrate-engaged singly capped (SC) proteasome (**b**), and the substrate-engaged doubly capped (DC) proteasomes (**c**) computed by AlphaCryo4D. 19 conformational states of the RP-CP subcomplex are marked on the energy landscape. **d**, Comparison of the state distribution of RP in the SC and DC proteasomes. **e**, State distribution matrix of the DC proteasome obtained using AlphaCryo4D on experimental data, colored by the particle numbers in the DC states, with the horizonal and vertical axes representing the states of two RPs bound to the same CP. **f**, State distribution matrix of the DC proteasome predicted by a control model assuming that the two RPs in the same DC proteasome are completely independent of each other or uncoupled. The model assumes the same total number of DC particles and the same probability of observing each RP state as experimentally measured. **g**, State distribution difference matrix of the DC proteasome by experimental results minus the model predictions assuming uncoupled two RPs, colored by the absolute particle number difference. **h**, Relative differences of the DC proteasome in unit of percentage. The relative difference is the ratio of absolute difference divided by the particle numbers from the model prediction in each DC state.

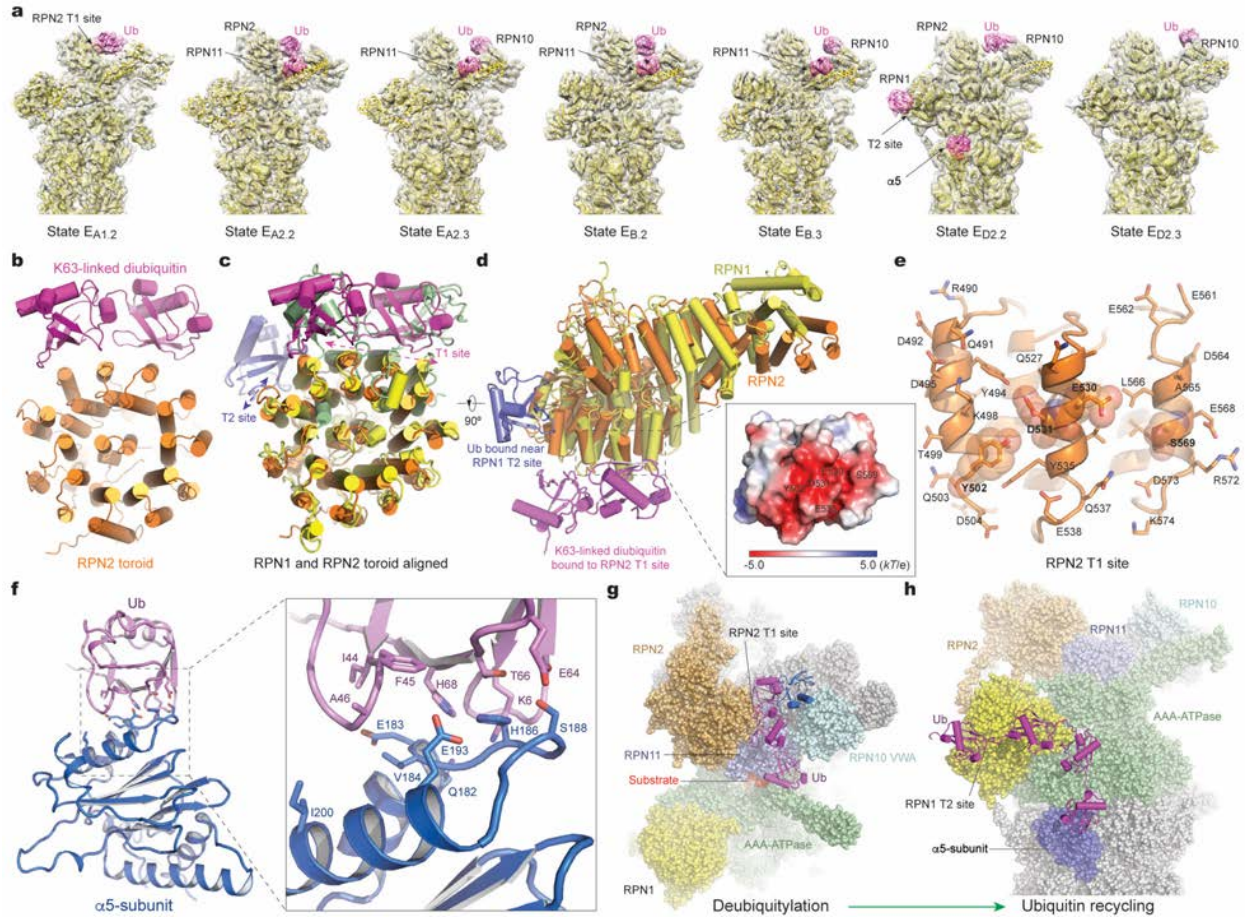

**Figure 4. Mapping of ubiquitin-binding sites on the human 26S proteasome**. **a**, Seven lowly populated transient states visualize various ubiquitin-binding sites on the RPN1, RPN2, RPN10 and α5 subunits, in addition to the RPN11 deubiquitylation site and previously observed RPT5 coiled-coil site[2]. For clarity of illustrating moderate-resolution ubiquitin (Ub) features, the cryo-EM maps of these states are low-pass filtered at 8 Å and shown in transparent surface representations superimposed with their corresponding atomic models in cartoon representations. The Ub densities are colored magenta, with the rest of the structures colored yellow. **b**, Structure of the RPN2 toroidal domain (orange) in complex with Lys63-linked diubiquitin (magenta) modelled from the cryo-EM maps of states $E_{A1.2}$, $E_{A2.2}$ and $E_{B.2}$, in cartoon representation. **c**, Structure-based alignment of the human RPN1 and RPN2 toroidal domains shown in color yellow and orange, respectively. The RPN1 T2 site-bound ubiquitin is shown as light blue. The NMR structure (PDB ID 2N3V) of the Lys48-linked dibuiquitin-bound yeast Rpn1 T1 site segment[11], shown as light green, is aligned with the human RPN1 structure and is superimposed for comparison. **d**, The entire human RPN1 and RPN2 subunit structures are aligned together and viewed from a perspective that is rotated 90° relative to the orientation in panel (**c**), showing that the two subunits are structurally homologous. Insert, electrostatic surface of the RPN2 T1 site showing its acidic nature. **e**, Closeup view of the RPN2 T1 site composed of three helices in carton representation with the side chains shown in stick representation. The key residues contacting ubiquitin are highlighted with transparent sphere representation. **f**, Structure of ubiquitin-bound α5 subunit shown in cartoon representation. The side chains of residues involved in the intermolecular interactions are shown in stick representation and are zoomed in on the right. **g**, A Lys63-linked triubiquitin chain model fitted from the cryo-EM maps of states $E_{A2.2}$ and $E_{B.2}$ is shown in purple cartoon representation and spans from the RPN2 T1 site to the RPN11 deubiquitylation site. The RPN10 VWA-bound monoubiquitin fitted in states $E_{A2.3}$ and $E_{B.3}$ is superimposed in blue cartoon representation. The proteasome is shown as space-filling sphere representations. **h**, A Lys63-linked tetraubiquitin chain model derived from the cryo-EM map of state $E_{D2.2}$. The two terminal ubiquitin molecules are fitted from their highest-resolution cryo-EM densities available at the RPN1 T2 and α5-subunit sites, whereas the middle two molecules are hypothetically modelled.

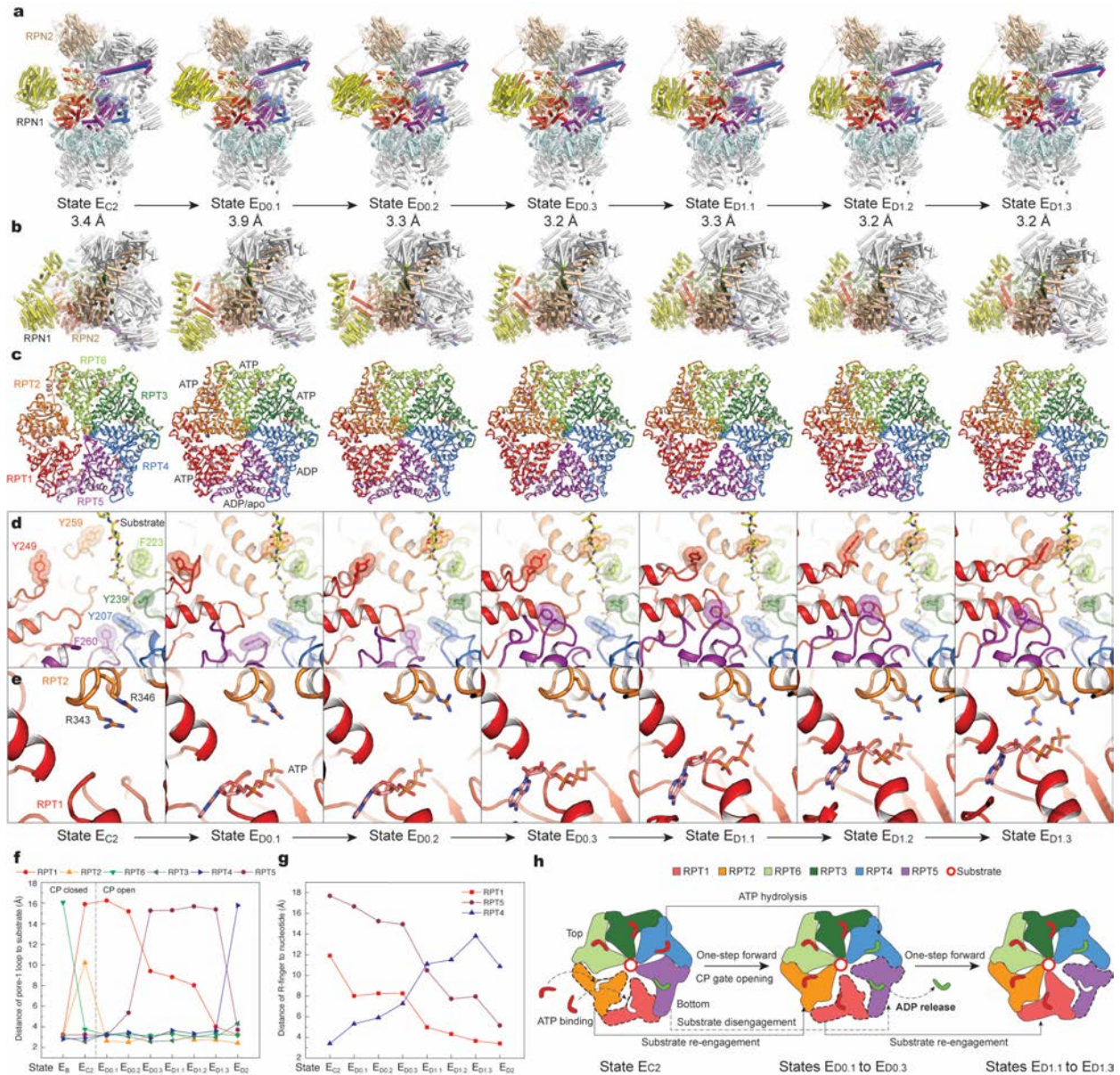

**Figure 5. Single-nucleotide exchange dynamics of the proteasomal AAA-ATPase motor during translocation initiation**. **a**, Side views of the substate-engaged RP-CP subcomplex in seven sequential conformers between states $E_{C2}$ and $E_{D1.3}$, as shown in cartoon representation and with only half CP shown. **b**, Top views of the seven sequential conformers, showing the overall lid rotation and the relative rotation between RPN1 and the rest of the RP structure. **c**, The substrate-bound AAA-ATPase structures of the seven conformers in cartoon representations. The substrates and nucleotides are shown in stick representations. **d**, Close-up views of the pore-1 loop interactions with the substrate (in yellow stick representation), with the aromatic residues in the pore-1 loops highlighted by transparent sphere representations. The seven snapshots show

that the RPT1 pore-1 loop is gradually moved to a position in direct contact with the substrate, whereas the RPT5 pore-1 loop is gradually moved away from the substrate and flipped up and out. **e**, Close-up views of the RPT1 nucleotide-binding sites in different states, showing its gradual closeup by the R-fingers from RPT2. **f**, Plots of distances of the pore-1 loops of all RPT subunits to the substrate measured in the atomic structures of 9 sequential conformational states. **g**, Plots of distances of R-finger to the nucleotides bound to three RPT subunits measured in the atomic structures of 8 sequential conformational states. **h**, Schematic illustrating the ATP hydrolysis reaction and nucleotide exchange associated with the intermediates during the transition from state $E_{C2}$ to $E_{D1.3}$.

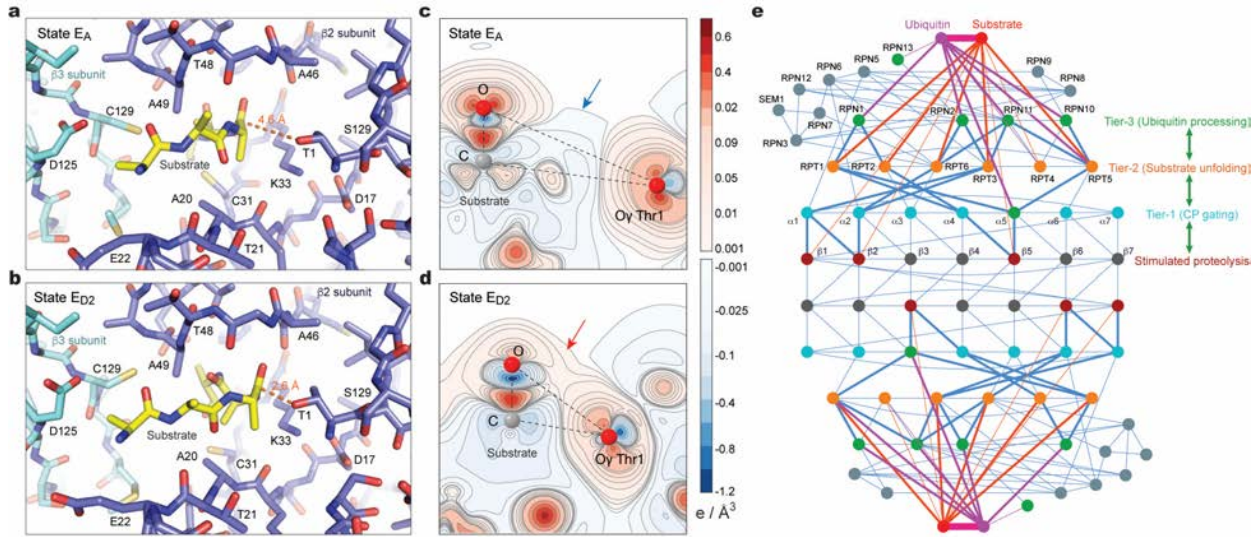

**Figure 6**. **Hierarchical allosteric regulation of proteasome activity**. **a** and **b**, Structure of the catalytic site with a substrate polypeptide bound at residue Thr1 in the β2 subunit of the CP in the closed-gate state $E_A$ at 2.7-Å resolution (**a**) and in the open-gate state $E_{D2}$ at 2.5-Å resolution (**b**). **c** and **d**, Charge density difference contour maps at the proteolytic active site of the β2 subunit computed by the quantum mechanical DFT based on the atomic structures of states $E_A$ (**c**) and $E_{D2}$ (**d**). The contours are plotted in the plane passing through the Thr1-Oγ atom and the nearest carbonyl carbon and oxygen atoms of the substrate. The interactions of the Thr1-Oγ atom and the carbonyl group lead to the nucleophilic attack as the first step of the proteolytic process. Due to the proximity shown in (**b**), a bonding interaction starts to form in the $E_{D2}$ states, indicated by the slight positive charge difference region (red arrow), whereas there is no such region formed in the $E_A$ state and the pair is well separated by a non-bonding charge depletion region (blue arrow). **e**, An interactome network diagram of the proteasome illustrating the grand allosteric regulation model and highlighting the key inter-subunit interaction pathways for proteasome autoregulation. The base subunits hosting ubiquitin-binding sites and the deubiquitinase RPN11 are shown as the green nodes; the AAA-ATPase subunits as the orange nodes; the α-type subunits as the cyan nodes; the catalytically active β1, β2 and β5 subunits as the crimson nodes. Steal blue lines represent inter-subunit interactions; purple lines represent the interactions with ubiquitin; and red lines represent substrate polypeptide interactions. Bold lines highlight the shortest pathways connecting the three tiers of regulatory subunits that are expected to propagate the allosteric regulation from ubiquitin interactions down to the six proteolytic

active sites in the CP. The network diagram is constructed according to all 64 structural models of the substrate-bound proteasome in this study.

**Methods**

**Overview of AlphaCryo4D framework**

The design of AlphaCryo4D follows a few key ideas and objectives. First, several previous studies have demonstrated the beneficial effects of energy landscape reconstitution of macromolecules from cryo-EM data[5,40,51-55]. The energy landscape is a statistical representation of the conformational space of a macromolecule and is the basis of the transition-state theory of reaction dynamics[56,57]. The minimum-energy path (MEP) on the energy landscape theoretically represents the most probable trajectory of conformational transitions and can inform the activation energy for chemical reactions[57,58]. We hypothesize that a high-quality reconstitution of energy landscape could be potentially used to improve 3D classification of cryo-EM data, which is a primary objective of AlphaCryo4D. If different conformers or continuous conformational changes can be sufficiently mapped and differentiated on the energy landscape, the energetic visualization of conformational continuum could then be used to discover previously missing conformers and achieve higher resolution for lowly populated, transient or nonequilibrium states.

Second, because the signal-to-noise ratio (SNR) of experimental cryo-EM data is often in the range of 0.005-0.05, most objective functions in image similarity measurement and machine learning algorithms tend to poorly perform or entirely fail. To date, no cryo-EM methods have included any implicit metrics or *en bloc* strategies for validating the 3D classification accuracy, although the final density map quality, resolution, key features and atomic modelling have been well used for explicit validation without knowing the exact 3D classification accuracy. To address this issue, inspired by the recent study on the robustness of dual objective functions in detecting weak signals[59], an integrative procedure called 'particle shuffling and voting' was devised in AlphaCryo4D to allow implicit cross-validation for 3D classification.

Third, conventional 3D classification methods in cryo-EM follow steps of hierarchical 3D clustering, alternating 2D and 3D classification, or combination of both[16,17,60]. Low-quality classes, no matter due to misclassification or misalignment, are then manually removed in the intermediate steps of hierarchical classification. This could result in considerable loss of potentially useful particles. The final results strongly depend on user expertise of inspecting intermediate classification steps and decision-making in class selection, which can be biased by user experience and subjectivity. Two limitations of such a procedure are: (1) a considerable portion of misclassified images in observed conformational states that limits their achievable resolution, and (2) missing conformational states of low populations that could be biologically important. To overcome these limitations, AlphaCryo4D was expected to simultaneously optimize the usage of all available particles and their 3D classification accuracy.

In practice, AlphaCryo4D consists of four major steps (Fig. 1a). First, all particles in the original dataset are aligned in a common frame of reference through a consensus 3D refinement in RELION[6]. All single-particle images are then randomly divided into $M+1$ groups of equal data sizes. In the step called '$M$-fold particle shuffling', one group of particles is taken out of the dataset to form a shuffled dataset. This procedure is repeated for $M + 1$ times, each time with a different particle group being left out, resulting in $M + 1$ shuffled datasets (Extended Data Fig. 1). Each shuffled dataset is subject to 3D volume bootstrapping separately and is clustered into tens to hundreds of 3D reconstructions through Bayesian clustering in RELION[6,18]. In total, thousands of volumes from all shuffled datasets are expected to be bootstrapped through these steps. Second, all bootstrapped volumes are learned by a 3D autoencoder made of a deep residual convolutional neural network in an unsupervised fashion[3,23] (Fig. 1b, Extended Data Table 1). A 3D feature map is extracted for each volume and is juxtaposed with the volume data for nonlinear dimensionality reduction by manifold embedding with the t-distributed stochastic neighbor embedding (t-SNE) algorithm[4]. Third, the learned manifold is used to compute an energy landscape via the Boltzmann relation[5]. A string method is used to search the MEP on the energy landscape[24,25]. The local energy minima or transition states connecting adjacent minimum-energy states can be defined as the centers of 3D clustering, with a circular range defined as cluster boundaries for subsequent particle voting. Fourth, because each particle is used $M$ times during volume bootstrapping, it is mapped to $M$ locations on the energy landscape. The mapping of each copy of the particle is called a 'vote'. By counting the number ($K$) of votes of the same particle being mapped within the same cluster boundary on the energy landscape, the reproducibility of manifold learning can be evaluated at single-particle level. Each particle is classified to the 3D cluster that receives more than $M/2$ votes of this particle. If none of the 3D clusters on the energy landscape receives more than $M/2$ votes of a given particle, the corresponding particle is voted out in the procedure and is excluded for further 3D reconstruction. As a result, each of the final 3D reconstructions includes structurally homogeneous particles that can potentially achieve higher resolution through dedicated cryo-EM structure refinement in RELION[6] or cryoSPARC[8].

**Data-processing workflow of AlphaCryo4D**

**Input**: Single-particle cryo-EM dataset after initial particle rejection of apparent junks.

**Output**: Energy landscape, MEP, 3D class assignment of each particle.

**Step 1**. Bootstrap a large number of 3D volumes through particle shuffling, consensus alignment and Bayesian clustering.

1. Split the particle dataset randomly to many sub-datasets, if necessary, in case of a large dataset (>150,000 particle images), for batch processing of particle shuffling and volume bootstrapping. Otherwise, skip this step if the dataset is small enough (<150,000).
2. For each sub-dataset, conduct a consensus alignment to generate initial parameters of Euler angles and translations in RELION.
3. Divide each sub-dataset into $M + 1$ groups, shuffle the sub-dataset $M + 1$ times and each time take a different group out of the shuffled sub-dataset, giving rise to $M + 1$ shuffled sub-datasets all with different collection of particles.
4. Conduct 3D Bayesian classification on all the $M + 1$ shuffled sub-datasets to generate hundreds of 3D volumes, making each particle to contribute to $M$ different volumes.
5. Execute steps (2) to (4) using the same initial model (low-pass filtered at 60-Å) for all sub-datasets.

**Step 2**. Extract deep features of all volume data with the 3D deep residual autoencoder.

6. Align all 3D volumes and adjust them to share a common frame of reference.
7. Initialize the hyper-parameters of the 3D autoencoder (Extended Data Table 1).
8. Train the neural network with the 3D volume data to minimize mean square error between the decoding layer and the input by the Adam algorithm of initial learning rate 0.01.
9. Extract the 3D feature maps of all volumes from the encoding layer.

**Step 3**. Embed the volume data to two-dimensional manifolds through the t-SNE algorithm and compute the energy landscape and find the MEP.

10. Calculate the pairwise similarities between volumes using their feature-map-expanded volume vectors, and randomly initialize the low-dimensional points.
11. Minimize the Kullback-Leibler divergence by the Momentum algorithm to generate 2D manifold embeddings with t-SNE.

12. Compute the energy landscape from the manifold using the Boltzmann relation.
13. Initialize searching of the MEP with a straight line between given starting and ending points.
14. Find the optimal MEP solution using the string method.

**Step 4**. Classify all particles through the energy-based particle-voting algorithm.

15. Sample the clustering centers along the MEP and calculate the recommended clustering radius.
16. Define the local energy minima as the 3D clustering centers and their corresponding cluster boundary for particle voting.
17. For each particle, cast a labelled vote for a 3D class when a volume containing one of the $M$ particle copies is located within the voting boundary.
18. Count the number of votes of each particle with respect to each 3D class and assign the particle to the 3D class that receives more than $M/2$ votes from this particle.
19. Refine each 3D density map separately to high resolution in RELION or cryoSPARC using particles classified into the same 3D classes.

**Particle shuffling for bootstrapping 3D volumes**

A key philosophy of the AlphaCryo4D design is to avoid subjective judgement on the particle quality and usability as long as it is not apparent junks like ice contaminants, carbon edges and other obvious impurities. Deep-learning-based particle picking in DeepEM[61], Torpaz[62] or other similarly performed software is favored for data preprocessing prior to AlphaCryo4D. To prepare particle datasets for AlphaCryo4D, an initial unsupervised 2D image classification and particle selection, preferentially conducted by the statistical manifold-learning-based algorithm in ROME[16], is necessary to ensure that no apparent junks are selected for further analysis and that the data have been collected under an optimal microscope alignment condition, such as optimized coma-free alignment. No particles should be discarded based on their structural appearance during this step if they are not apparent junks. Any additional 3D classification should be avoided to pre-maturely reject particles prior to particle shuffling and volume bootstrapping in the first step of AlphaCryo4D processing. Pre-maturely rejecting true particles via any 2D and 3D classification is expected to introduce subjective bias and to impair the native conformational continuity and statistical integrity intrinsically existing in the dataset.

In raw cryo-EM data, 2D transmission images of biological macromolecules suffer from extremely heavy background noise, due to the use of low electron dose to avoid radiation damage. To tackle the conformational heterogeneity of the macromolecule sample of interest in the presence of heavy image noise, the particle shuffling and volume bootstrapping procedure was devised to incorporate the Bayesian or maximum-likelihood-based 3D clustering in RELION[6]. To enable the particle-voting algorithm in the late stage of AlphaCryo4D, each particle is reused $M$ times during particle shuffling to bootstrap a large number of 3D volumes (Extended Data Fig. 1a). First, all particle images are aligned to the same frame of reference in a consensus 3D reconstruction and refinement in RELION[6] or ROME[16] to obtain the initial alignment parameters of three Euler angles and two translational shifts. Optimization for alignment accuracy should be pursued to avoid the error propagation to the subsequent steps in AlphaCryo4D. For a dataset with both compositional and conformational heterogeneity, coarsely classifying the dataset to a few 3D classes during initial image alignment, limiting the 3D alignment to a moderate resolution like 10 Å or 15 Å during global orientational search, progressing to small enough angular steps in the final stage of consensus refinement, may be optionally practiced in order to optimize the initial 3D alignment. Failure of initial alignment of particles would lead to failure of all subsequent AlphaCryo4D analysis.

Next, based on the results of consensus alignment, in the particle-shuffling step, all particles were divided into $M + 1$ groups, and $M$ was set to an odd number of at least 3. Then the whole dataset was shuffled $M + 1$ times by removing a different group out of the dataset each time. Each shuffled dataset is classified into a large number ($B$) of 3D volumes, often tens to hundreds, by the maximum-likelihood 3D classification algorithm without further image alignment in RELION (with 'skip-align' option turned on). This is necessary because the alignment accuracy often degrades when the sizes of 3D classes decrease considerably. This step is repeated $M + 1$ times, each time on a shuffled dataset missing a different group among the $M + 1$ groups. Due to the effect of $M$-fold particle shuffling, the outcome of this entire process is expected to bootstrap up to thousands of 3D volumes in total (i.e., $B(M +1) > 1000$). Each particle is used and contributed to the 3D reconstructions of $M$ volumes, which prepare it for the particle-voting algorithm to evaluate the robustness and reproducibility of each particle with respect to the eventual 3D classification.

For processing a large dataset including millions of single-particle images, it becomes infeasible even for a modern high-performance computing system to do the consensus alignment by including all particles once in a single run due to limitation of supercomputer memory and the scalability of the alignment software such. To tackle this issue, the whole dataset is randomly split into a number ($D$) of sub-datasets for batch processing, with each sub-dataset including about one to two hundred thousand particles, depending on the scale of available supercomputing system. In this case, the initial reference should be used for the consensus alignment of different sub-datasets to minimize volume alignment errors in the later step. The total number of resulting bootstrapped volumes becomes $BD(M+1) > 1000$. This strategy can substantially reduce the supercomputer memory pressure and requirement. In each sub-dataset, all particles were divided into $M + 1$ groups and subject to the particle shuffling and volume bootstrapping procedure as described above. To balance the computational cost and algorithmic efficiency, we used $M = 3$ in the entire data processing workflow involved in the current study, implying that every particle was used three times for subsequent computation before particle voting. But a higher $M$ value might be beneficial for smaller dataset or lower SNR, whose effects are not investigated in the present study due to the limit of computational resources and feasibility.

**Deep residual autoencoder for 3D feature extraction**

The bootstrapped volumes may still suffer from reconstruction noises and errors due to variation of particle number, misclassification of conformers and limited alignment accuracy. Thus, we hypothesize that unsupervised deep learning could help capture the key features of structural variations in the bootstrapped volumes and potentially enhance the quality of subsequently reconstituted energy landscapes for improved 3D classification. To this end, a 3D autoencoder was constructed using a deep Fully Convolutional Network (FCN) composed of residual learning blocks[3,23]. The architecture of the 3D autoencoder consists of the encoder and the decoder, which are denoted as $\mathcal{E}$ and $\mathcal{D}$, respectively (Fig. 1b, Extended Data Table 1). The relation between the output $\boldsymbol{y}$ and the input $\boldsymbol{x}$ of the network can be expressed as:

$$\boldsymbol{y} = \mathcal{D}(\mathcal{E}(\boldsymbol{x})), \quad (1)$$

in which $\boldsymbol{x}$ is the input 3D density volume with the size of $N^3$, where $N$ is the box size of the density map in pixel units. For reconstruction of the 3D volumes and further optimization, the decoding output $\boldsymbol{y}$ should be in the same size and range with the input data $\boldsymbol{x}$. In this way, the framework of FCN is established to restore the input volume, using the sigmoid function $S(x) =$ $\frac{1}{1+\exp{(-x)}}$ as the activation function of the decoding layer to normalize the value of $y$ into the range (0, 1). Meanwhile, all 3D density maps $x$ should be preprocessed with the equation (2) before input to the deep neural network:

$$x_{ijk} := \frac{x_{ijk} - x_{min}}{x_{max} - x_{min}}, \quad i, j, k = 1, 2, \ldots, N. \quad (2)$$

where $x_{min}$ and $x_{max}$ are, respectively, the minimum and minimax value in all $x_{ijk}$.

The distance between the decoded maps and the input volumes can be used for constructing the loss function to train the 3D kernels and bias of the networks. The value distribution of the encoded 3D feature maps $z = \mathcal{E}(x)$ is expected to be an abstract, numerical representation of the underlying structures in the volume data, which may not necessarily have any intuitive real-space physical meanings. The neural network is capable of extracting such abstract information in the prediction step, with no restriction on the feature maps $z$ in the expression of training loss. The loss function is then formulated as:

$$L(\boldsymbol{\theta}; \boldsymbol{x}, \boldsymbol{y}) = \frac{1}{N^3} \sum_{i,j,k=1}^{N} \|x_{ijk} - y_{ijk}\|^2 + \lambda \|\boldsymbol{\theta}\|^2, \quad (3)$$

where $\boldsymbol{\theta}$ denotes the weights and bias of the network, and $\lambda$ is L2 norm regularization coefficient. As the feature of the complex structure is difficult to be learned from the 3D volume data, the value of $\lambda$ is generally set to 0 to focus on the first term of the expression (3) unless overfitting arises.

To improve the learning capacity of the 3D autoencoder, residual learning blocks containing 3D convolutional and transposed convolutional layers are employed in the encoder and decoder, respectively. In each residual block, a convolutional layer followed by a Batch Normalization[63] (BN) layer and activation layer appears twice as a basic mapping, which is added with the input to generate the output. The mathematical expression of the $l$th block can be shown as:

$$\begin{cases} \boldsymbol{y}_l = \mathcal{F}(\boldsymbol{x}_l; \boldsymbol{\theta}_l) + \boldsymbol{x}_l \\ \mathcal{F}(\boldsymbol{x}_l; \boldsymbol{\theta}_l) = \mathcal{C}(\mathcal{C}(\boldsymbol{x}_l)), \quad (4) \\ \mathcal{C}(\boldsymbol{x}_l) = a(b(c(\boldsymbol{x}_l))) \end{cases}$$

where $\boldsymbol{x}_l$ and $\boldsymbol{y}_l$ represent the input and the output of the block, respectively. $\mathcal{F}(x_l; \boldsymbol{\theta}_l)$ denotes the basic mapping of the $l$th block parameterized by $\boldsymbol{\theta}_l$, and $\mathcal{C}(\boldsymbol{x}_l)$ is the sequential operation of convolution or transposed convolution $c$, BN $b$ and activation $a$. The rectified linear unit (ReLU) function is used in all the activation layers but the last one. In addition, the function (4) demands that the output of mapping function $\mathcal{F}(x_l; \theta_l)$ must have the same dimension as the input $x_l$. If this is not the case, the input $x_l$ must be rescaled along with $\mathcal{F}(x_l; \theta_l)$ using a convolutional or transposed convolutional transformation $c'(x_l)$ with an appropriate stride value, the parameters of which can be updated in the training step.

To analyze a large number of volumes, we tuned carefully the training of the 3D deep residual autoencoder to obtain suitable kernels and bias. First, parallel computation with multiple GPUs has been implemented to reduce the training time. Then the parameters of the network are optimized by the stochastic gradient descent Adam (Adaptive moment estimation) algorithm, in which the gradients of the objective function $L(\theta; x, y)$ with respect to the parameters $\theta$ can be calculated by the chain rule. Moreover, the learning rate is reduced to one tenth when the loss function does not decrease in three epochs based on the initial value of 0.01. After trained about 50 epochs, the best model is picked to execute the task of structural feature extraction. Using the unsupervised 3D autoencoder, the feature maps $z = \mathcal{E}(x)$ encoding the structural discrepancy among the 3D volume data can be extracted automatically without any human intervention.

**Manifold embedding of energy landscape**

To prepare for the energy landscape reconstitution, each bootstrapped 3D volume was juxtaposed with its 3D feature map learned by the 3D autoencoder to form an expanded higher-dimensional data point. All the expanded data points were then embedded onto a low-dimensional manifold via the t-SNE algorithm by preserving the geodesic relationships among all high-dimensional data[4]. During manifold embedding, it is assumed that the pairwise similarities in the high dimensional data space and low dimensional latent space follow a Gaussian distribution and a Student's t-distribution, respectively, which can be formulated as:

$$\begin{cases} p_{ij} = \dfrac{\exp(-\|x_i - x_j\|^2 / 2\sigma_i^2)}{\sum_{i \neq j} \exp(-\|x_i - x_j\|^2 / 2\sigma_i^2)}, & i \neq j \\ q_{ij} = \dfrac{(1 + \|y_i - y_j\|^2)^{-1}}{\sum_{i \neq j}(1 + \|y_i - y_j\|^2)^{-1}}, & i \neq j \end{cases} \quad (5)$$

where $p_{ij}$ and $q_{ij}$ represent the similarity distributions in the high and low dimensional spaces, $x_k$ and $y_k$ ($k = i, j$) are the data points of the high and low dimensional spaces, respectively. The parameter $\sigma^2$ is the variance of the Gaussian distribution. In addition, $p_{ii}$ and $q_{ii}$ are both set to zero to satisfy the constraint of symmetry.

To find the value $y_k$ of each data point, an objective function measuring the distance between the similarity distribution $p_{ij}$ and $q_{ij}$ had to be well defined. Here the relative entropy, also called the Kullback-Leibler (KL) divergence,

$$KL(P \parallel Q) = \sum_{i,j} p_{ij} \log \frac{p_{ij}}{q_{ij}}, \quad (6)$$

was employed and minimized by the gradient descent algorithm with the momentum method[64].

The idea of using juxtaposed data composed of both volumes and their feature maps for manifold embedding is similar to the design philosophy of deep residual learning, in which the input and feature output of a residual learning block are added together to be used as input of the next residual learning block. Such a design has been demonstrated to improve learning accuracy and reduce the performance degradation issues when the neural network goes much deeper[3,23]. Although the 3D feature maps or the bootstrapped volumes alone can be used for manifold embedding, both appear to be inferior in 3D classification accuracy (Extended Data Fig. 5). The juxtaposed format of input data for manifold learning is potentially beneficial to the applications in those challenging scenarios, such as visualizing a reversibly bound small protein like ubiquitin (~8.6 kDa) of low occupancy by the focused AlphaCryo4D classification (see below).

After the manifold embedding by t-SNR, each 3D volume is mapped to a low-dimensional data point in the learned manifold. The coordinate system, in which the low-dimensional representation of the manifold is embedded, is used for reconstructing energy landscape. The difference in the Gibbs free energy $\Delta G$ between two states with particle numbers of $N_i$ and $N_j$ is defined by the Boltzmann relation $N_i/N_j = \exp(-\Delta G/k_B T)$. Thus, the free energy of each volume can be estimated using its corresponding particle number:

$$\Delta\Delta G_i = -k_B T \ln \frac{N_i}{\sum_i N_i}, \quad (7)$$

where $\Delta\Delta G_i$ denotes the free energy difference of the data point with the particle number of $N_i$ against a common reference energy level, $k_B$ is the Boltzmann constant and $T$ is the temperature in Kelvin. The energy landscape was plotted by interpolation of the free energy difference in areas with sparse data. We suggest that linear interpolation is used for the energy landscape with loosely sampled areas to avoid overfitting by polynomial or quadratic interpolation. For densely sampled energy landscape, polynomial interpolation could give rise to smoother energy landscape that is easier to be tackled by the string method for MEP solution (see below).

The coordinate system of the embedded manifold output by t-SNE is inherited by the reconstitution of energy landscape as reaction coordinates. They do not have intuitive physical meaning of length scale in the real space. However, they can be viewed as transformed, rescaled, reprojected coordinates from the real-space reaction coordinates along which the most prominent structural changes can be observed. Alternatively, they can be understood as being similar to transformed, reprojected principal components (PCs) in principle component analysis (PCA). In the case of the 26S proteasome, the two most prominent motions are the rotation of the lid relative to the base and the intrinsic motion of AAA-ATPase motor[2], which can be approximately projected to two reaction coordinates (Fig. 3a-c). In real space, both motions are notably complex and in fact are characterized by a considerable number of degrees of freedom, which are partially defined by the solved conformers.

**String method for finding minimum energy path (MEP)**

The string method is an effective algorithm to find the MEP on the potential energy surface[24]. To extract the dynamic information implicated in the experimental energy landscape, an improved and simplified version of the string method has been previously developed[25]. Along the MEP on the energy landscape, the local minima of interest could be defined as 3D clustering centers to guide the particle-voting algorithm for 3D classification to generate high-resolution cryo-EM density maps (Extended Data Fig. 1b). The objective of the MEP identification in energy barrier-crossing events lies in finding a curve $\gamma$ having the same tangent direction as the gradient of energy surface $\nabla G$. It can be expressed as:

$$(\nabla G)^\perp(\gamma) = 0, \quad (8)$$

where $(\nabla G)^\perp$ denotes the component of $\nabla G$ perpendicular to the path $\gamma$. To optimize the objective function (4), two computational steps, previously named evolution of the images and reparameterization of the string, are iterated until convergence within a given precision threshold.

*Evolution of the images.* After initialization with the starting and ending points, the positions of interval images were updated according to gradient of the free energy at the $t$th iteration:

$$\varphi_i^{*(t)} = \varphi_i^{(t-1)} - h\nabla G\left(\varphi_i^{(t-1)}\right), \quad (9)$$

with $\varphi_i^{(t)}$ ($i = 0,1,\ldots,N$) being the $i$th intermediate image at the $t$th iteration ($*$ denoting the temporary values), and $h$ the learning rate.

*Reparameterization of the string.* The values of positions $\varphi_i^{(t)}$ ($i = 0,1,\ldots,N$) were interpolated onto a uniform mesh with the constant number of points. Prior to interpolation, the normalized length $\alpha_i^{*(t)}$ ($i = 0,1,\ldots,N$) along the path was calculated as:

$$\alpha_0^{*(t)} = 0, \quad \alpha_i^{*(t)} = \alpha_{i-1}^{*(t)} + \frac{\left\|\varphi_i^{*(t)} - \varphi_{i-1}^{*(t)}\right\|}{\sum_{i=1}^{N}\left\|\varphi_i^{*(t)} - \varphi_{i-1}^{*(t)}\right\|}, \quad i = 1,2,\ldots,N. \quad (10)$$

Given a set of data points $(\alpha_i^{*(t)}, \varphi_i^{*(t)})$, the linear interpolation function was next used to generate the new values of positions $\varphi_i^{(t)}$ ($i = 0,1,\ldots,N$) at the uniform grid points $\alpha_i^{(t)}$ ($i = 0,1,\ldots,N$). The iteration was terminated when the relative difference $\sum_{i=0}^{N}\left\|\varphi_i^{(t)} - \varphi_i^{(t-1)}\right\|^2/N$ became small enough.

**Energy-based particle voting algorithm**

The particle-voting algorithm was designed to conduct 3D classification, particle quality control, reproducibility test and particle selection in an integrative manner. The particle-voting algorithm mainly involves two steps (Extended Data Fig. 1b). First, we count the number of votes for each particle mapped within the voting boundaries of all 3D clusters on the energy landscape. One vote is rigorously mapped to one copy of the particle used in reconstructing a 3D volume and to no more than one 3D cluster on the energy landscape where the corresponding volume is located. Thus, each particle can have $M$ votes casted for no more than $M$ 3D clusters. If the vote is mapped outside of any 3D cluster boundary, it becomes an 'empty vote' with no cluster label. Each non-empty vote is thus labeled for both its particle identify and corresponding cluster identity. For each pair of particle and cluster, we compute the total number ($K$) of votes that the cluster receives from the same particle. Each particle is then assigned and classified to the 3D cluster that receives $K > M/2$ votes from this particle (Extended Data Fig. 1b). Note that after particle voting, each particle is assigned no more than once to a 3D class, with its redundant particle copies removed from this class. This strategy only retains the particles that can reproducibly vote for a 3D cluster corresponding to a homogeneous conformation, while abandoning those non-reproducible particles with divergent, inconsistent votes.

Because the particle-voting algorithm imposes strong constraints on the numeric performance of particles in deep manifold learning, it could lead to particle number insufficiency in the cases of smaller but potentially interested 3D classes. To remedy this limitation, an alternative, distance-based classification algorithm was devised to replace the particle-voting algorithm when there are not enough particles to gain the advantage of particle voting (Extended Data Fig. 1c). In this method, the distances of all $M$ copies of each particle to all 3D cluster centers on the energy landscape are measured and ranked. Then, the particle is classified to the 3D cluster of the shortest distance. A threshold could also be manually preset to remove particles that are too far away from any of the cluster centers. The distance-based classification method can keep more particles, but it ignores the potential issue of irreproducibility of low-quality particles. Thus, it is proven to be less accurate in 3D classification (Extended Data Fig. 5m-o). In other words, it trades off the classification accuracy and class homogeneity to gain more particles, which is expected to be potentially useful for small datasets or small classes. By contrast, the energy-based particle-voting algorithm imposes a more stringent constraint to select particles of high reproducibility during classification, resulting in higher quality and homogeneity in the classified particles, which is superior to the distance-based classification method (Extended Data Fig. 5s-u).

**Practical consideration of model parameters and network hyperparameters**

**Particle shuffling parameters**. The parameter $M$ determines the degree of implicit cross-validation of classification reproducibility, as well as the sampling densities of energy landscape. To establish reproducibility for 3D classification, $M$ should be no less than 3. The variation of $M$ is not supposed to change the energy landscape because it multiplies on both denominator and enumerator in equation (7) and is cancelled in the ratio of particle densities in computing the free-energy differences through the Boltzmann relation. Increasing $M$ value will allow more volumes to be bootstrapped, which then leads to a higher sampling density in computing the manifold of energy landscape and potentially enhances its reconstitution quality. The default classification threshold $M/2$ is considered to be minimal for verifying the reproducibility of 3D classification. However, a higher threshold, such as $2M/3$, will give rise to more stringent criteria in cross-validation, with a tradeoff of voting out more particles and of higher computational costs. The particle voting algorithm does not entirely eradicate misclassified particles. However, increasing either $M$ or the classification threshold could theoretically have similar impact and improve the conformational homogeneity, because it gets less probable for a particle being misclassified to the same cluster for more times.

Several considerations may be applied to the choice of *M* and to set up particle shuffling and volume bootstrapping. First, to obtain a high-quality energy landscape, a thousand or more bootstrapped volumes are expected for datasets of more than 1-2 million particles. Second, the average particle number per volume is expected to be no less than 5000 or more to ensure that majority of volumes include sufficient particles for quality reconstructions. The expected particle number per volume must increase if the average SNR per particle is decreased. It may need to reach 10000-20000 or more for small proteins or lower SNR datasets. Third, for a dataset of a moderate size, *M* can be adjusted to a higher value to mitigate the lack of image data for bootstrap.

**Autoencoder hyperparameters**. The recommended hyperparameters for the deep residual autoencoder architecture are provided in Extended Data Table 1. Although we expand the residual network (ResNet) into 3D, we keep the original design rules of ResNet for constructing the 3D autoencoder[3,23]. The first and last convolutional layers have 5 × 5 × 5 filters (kernels). The remaining convolutional layers have 3 × 3 × 3 filters. Because the cubic filter in convolutional layer is computationally expensive, only one 3D filter is used in each of the last three convolutional layers in the encoder and of the first two and last transposed convolutional layers in the decoder. To accommodate a large volume input, two filters are used in the first three convolutional layers and in the third and fourth transposed convolutional layers. Downsampling is directly performed by convolutional layers with a stride of 2. The output dimension of encoding layers is set to be a quarter of the input dimension, in order to avoid over-compressing the feature maps. We employed six 3D convolutional layers in the encoder and five transposed convolutional layers in the decoder based on the expected tradeoff between the learning accuracy and training cost, which is roughly comparable to a 2D ResNet with more than 200 layers with respect to the training cost.

**String method parameters**. The string method of searching a rational MEP on the energy landscape can only guarantee the solution of local optimum and is not capable of ensuring a solution of global optimum[25]. The outcome of the string method depends on the initialization of the starting and ending points of the MEP on the energy landscape. If there are too many local minima to sample along a long MEP, the string method could retrieve a MEP solution that partly misses some local energy minima by going off the pathway. In this case, the search of an expected MEP solution can be divided into several segments of shorter MEPs connecting one another, with each MEP being defined by a closer pair of starting and ending points, which travels through a smaller number of energy minima. Another parameter affecting the MEP solution is the step size being set to explore the MEP on energy landscapes. Although a default value of 0.1 empirically recommended may work in many cases, it may be helpful to tune the step size according to the quality of the energy landscape. The value of step size can be decreased if the computed path runs out of the energy landscape and can be increased if the path updates too slowly during iterative searching by the string method.

**Blind assessments with simulated datasets**

Three simulated large datasets with the SNRs of 0.05, 0.01 and 0.005, each including 2-million particle images, were employed to benchmark AlphaCryo4D and to compare its performance with alternative methods. For each synthetic dataset, the particles were computationally simulated by projecting the 20 3D density maps calculated from 20 hypothetical atomic models emulating continuous rotation of the NLRP3 inflammasome protein. The 20 atomic models were interpolated between the inactive NLRP3 structure and its hypothetical active state, which was generated through homology modeling using the activated NLRC4 structure[26]. The 20 atomic models represent sequential intermediate conformations during a continuous rotation in its NATCH domain against its LRR domain over an angular range of 90°. Each conformation is thus rotated 4.5° over its immediate predecessor in the conformational continuum sequence. 100,000 simulated particle images per conformational state were generated with random defocus values in the range of -0.5 to -3.0 μm, resulting in 2 million particles for each dataset of a given SNR. The pixel size of the simulated image was set to the same as the pixel size (0.84 Å) of the real experimental dataset of NLRP3-NEK7 complex[26]. To emulate realistic circumstances in cryo-EM imaging, Gaussian noises, random Euler angles covering half a sphere and random in-plane translational shifts from -5.0 to 5.0 pixels were then applied to every particle image.

   Each simulated heterogeneous NLRP3 dataset was analyzed separately by AlphaCryo4D and used to characterize the performance and robustness of AlphaCryo4D against the variation of SNRs. In the step of particle shuffling and volume bootstrapping, 2,000,000 particles in the dataset of any given SNR were divided randomly into 10 sub-datasets for batch processing. The orientation of each particle was determined in the initial 3D consensus alignment in RELION, which did not change in the subsequent 3D classification. In this step, the maximum number of iterations of the 3D alignment was set up as 30, with the initial reference low-pass filtered to 60 Å. 3-fold particle shuffling (indicated as × 3 below) was conducted on each sub-dataset for volume bootstrapping. The first round of maximum-likelihood 3D classification divided the input shuffled particle sub-dataset into 5 classes, each of these classes were then further classified into 8 classes. This procedure was separately executed on all shuffled particle sub-datasets. The particle shuffling and volume bootstrapping generated 1,372, 1,489, and 1,587 volumes by the datasets with SNRs of 0.05, 0.01 and 0.005, respectively. These volume data were used as inputs for deep residual autoencoder to compute low-dimensional manifolds and energy landscapes (Fig. 1d-f, Extended Data Fig. 2a, b). After searching the MEP on the energy landscapes by the string method, 20 cluster centers along the MEP were defined by the local energy minima along the MEP by approximately equal geodesic distance between adjacent minima, which represent potentially different conformations of the molecule (Extended Data Fig. 2a-c). The particle-voting algorithm was applied in every cluster to determine the final particle sets for all 3D classes. For the purpose of validation of the methodology and investigation of 3D classification improvement, we labeled each bootstrapped 3D volume with the conformational state that held the maximum proportion of particles in the class and computed its 3D classification precision as the ratio of the particle number belonging to the labelled class versus the total particle number in the volume (Figs. 1d, 2a, Extended Data Figs. 2a, b, 3a).

**Analysis of algorithm mechanism**

To understand how the 3D classification accuracy is improved by AlphaCryo4D, we analyzed the algorithmic mechanism by a number of control experiments (Extended Data Figs. 2d-g, 5). In total, we conducted 24 conditional control tests using 10 variations of algorithmic design by removing or replacing certain components of AlphaCryo4D. First, by removing the entire component of deep manifold learning, the distributions of high-precision 3D classes obtained by $M$-fold particle shuffling and voting alone drop by ~60% relative to that from complete AlphaCryo4D processing (Extended Data Fig. 5d-f). In these control experiments, the particle voting was achieved through counting the votes of a particle against the same conformers classified by RELION via computing the Intersection-over-Union (IoU) metric after the step of volume bootstrapping. Similarly, by removing the 3D deep residual autoencoder but keeping the manifold learning by t-SNE for energy landscape reconstitution, the distributions of high-precision 3D classes is reduced by ~15% at the SNR of 0.005 (Extended Data Fig. 5p-r). The effect of accuracy degradation is less prominent at a higher SNR (0.05 or 0.01), indicating that the unsupervised feature learning using deep residual autoencoder promotes the tolerance of the algorithm against higher noise level in the data. Together, these control tests suggest that deep manifold learning plays a crucial role in improving 3D classification accuracy with low SNR data.

Next, we replaced t-SNE with four other algorithms in manifold learning step, including two classic linear dimensionality reduction techniques, principal component analysis (PCA)[65] and multidimensional scaling (MDS)[66], and two nonlinear dimensionality reduction algorithms, isometric mapping (Isomap)[67] and locally linear embedding (LLE)[68]. We applied the four algorithms to reduce the dimensionality of the same sets of bootstrapped volume data. Although being capable of differentiating a portion of ground-truth conformers, the resulting 2D mappings by the four techniques exhibit considerable overlap between adjacent conformers and are incapable of distinguishing all 20 conformers of ground truth, thus inevitably missing many conformers (Extended Data Fig. 2d-g). The PCA and Isomap approximately missed half of the ground-truth conformers, whereas MDS and LLE both missed around 70% of the ground-truth conformers. The inferior performance of these techniques is consistent with the original control experiments conducted by the t-SNE developers[4].

Further, we conducted control experiments to evaluate the impact of particle voting on the 3D classification accuracy by replacing the particle voting component with a clustering strategy that directly classifies each particle to the cluster of the nearest clustering center (Extended Data Fig. 1c). In this case, the distributions of high-precision 3D classes are reduced by ~15-30% (Extended Data Fig. 5m-o). The reduction is more prominent at the lower SNR. This indicates that the particle voting considerably improves the 3D classification accuracy but is less impactful than the component of deep manifold learning.

Last, by tracking the statistics of classification precisions step by step, we evaluated how the 3D classification accuracy is improved over the intermediate steps of AlphaCryo4D (Extended

Data Fig. 5g-l). We found that the steps of particle shuffling, defining cluster boundaries on the energy landscapes and energy-based particle voting contribute to ~16%, ~20% and ~40% improvements in the distributions of high-precision 3D classes, respectively. Taken together, these results indicate that deep manifold learning is a crucial component and is further amplified in performance by energy-based particle voting. The two components appear to synergistically improve the 3D classification accuracy. Neither deep manifold learning nor particle voting alone is sufficient to achieve the present level of 3D classification accuracy.

**Comparisons with alternative methods**

Using 3D classification precision as a benchmark indicator, the performance of AlphaCryo4D was compared with several other methods: (1) the conventional maximum-likelihood-based 3D (ML3D) classification in RELION[6,7,17] with and without a hierarchical strategy, (2) the conventional heterogeneous reconstruction in cryoSPARC[8], (3) the 3DVA algorithm in cryoSPARC[21] and (4) the deep generative model-based cryoDRGN[22]. A total of 18 comparative tests by these alternative methods have been conducted blindly using the three simulated datasets, which includes 6-million images in total. In all our comparative tests on AlphaCryo4D and the alternative methods, the ground-truth information of particle orientations, in-plane translations and conformational identities were completely removed from the methods being tested and were not used for any steps of data processing or training. The ground-truth conformational identities of particles were only used as validation references to compute the 3D classification precisions of the blind testing results.

For the tests using ML3D in RELION, we classified all particles directly into 20 classes and hierarchically into 4 × 5 classes, which first divided the dataset into 4 classes, with each class further classified into five sub-classes (Extended Data Fig. 3). For testing the conventional discrete ab initio heterogeneous reconstructions in cryoSPARC, each synthetic dataset was directly classified into 20 classes without providing any low-resolution reference model (Extended Data Fig. 3). For comparison with the 3DVA algorithm in cryoSPARC, we first did the blind consensus alignment of the entire dataset to find the orientation of each particle. Then the alignment and the mask generated from the consensus reconstruction were used as inputs into the 3DVA calculation, with the number of orthogonal principal modes being set to 2 or 3 in 3D classification. The 3D variability display module in the cluster mode was used to analyze the results of 3D classification. For blind tests with cryoDRGN, a default 8D latent variable model was trained for 25 epochs. The encoder and decoder architectures were 256 × 3, as recommend by the cryoDRGN authors[22]. The particle alignment parameters prior to cryoDRGN training were obtained by the same blind consensus refinement in RELION used for other parallel tests. The metadata of 3D classification precisions as well as the 3D density maps from all the algorithms applied on the three simulated datasets were collected to conduct the statistical analysis (Fig. 2 and Extended Data Fig. 3).

The benchmarking results show the similar performance in 3D classification accuracy by ML3D in RELION and by heterogeneous reconstruction in cryoSPARC (Extended Data Fig. 3d-f). The averages of 3D classification precisions from these methods are all in the range of 0.23-0.34. The recently developed 3DVA algorithm in cryoSPARC represents an improvement over the conventional methods, with only 2-4 missing conformers and the average classification precisions reaching 0.36-0.55 when using 2 principal components (PCs) for 3D classification (Extended Data Fig. 3b). 3DVA using 3 PCs performed slightly better than using 2 PCs, but still missed at least two conformers, with the average classification precisions still below 0.6 in all three SNRs (Extended Data Fig. 3g). At the SNR of 0.05, cryoDRGN also outperformed the conventional 3D classification in RELION and cryoSPARC and was approximately on par with 3DVA with 2 PCs (Extended Data Fig. 3c). Its performance degraded with decreasing the SNR and became the worst among all tested methods at the SNR of 0.005 (Extended Data Fig. 3b, g). The lower SNR also reduced the average classification precision of 3DVA to 0.36-0.42. But it did not have obvious effects for ML3D in RELION. These comparisons indicate that cryoDRGN is most sensitive against the variation of SNRs among the tested methods at least in the present cases. By missing 2-7 conformers of ground truth, these methods failed in reconstruction of the full conformational continuum of ground truth (Extended Data Fig. 3b-g). The underlying cause of low classification accuracy by 3DVA and cryoDRGN can be understood by visualizing the particle distributions in feature space, which show considerable overlaps between the distributions of adjacent conformers (Extended Data Fig. h-j).

By contrast, AlphaCryo4D succeeded in these difficult tests by reconstructing the full conformational continuum of ground truth at all three SNRs (Fig. 1c, Extended Data Figs. 2, 3a, 3k, 3o, 3p, 4a). It achieved an average classification precision of 0.82-0.83 at the SNRs of 0.01 and 0.05, which are more than three times those achieved by the two conventional methods (ML3D in RELION and discrete heterogeneous reconstruction in cryoSPARC) and are 1.5-3 times those achieved by the recently proposed 3DVA and cryoDRGN methods. With decreasing the SNR to 0.005, the average classification precision of AlphaCryo4D was reduced to 0.65 and still outperformed the second-best results (0.56-0.59) obtained from 3DVA or cryoDRGN at the higher SNR of 0.05 (Extended Data Fig. 3a-c, g).

**Computational costs.** Although the computational cost of AlpahCryo4D is generally higher than the conventional approach, it does not appear to increase drastically and likely falls in an affordable range, while reducing the average cost of computation per conformational state. In a nutshell, we can have a brief comparison of the computational efficiency on the simulated 2-million image dataset with an SNR of 0.01. In the step of 3D data bootstrapping, we split the dataset into 20 subsets, which contained 100,000 particles each. The 3D consensus alignment of all 2,000,000 particles cost about 75 hours using 8 Tesla V100 GPUs interconnected with the high-speed NVLink data bridge in a NVIDIA DGX-1 supercomputing system. Within each subset, the 3D Bayesian classification for one leave-one-group-out dataset cost about 2.5 hours using 320 CPU cores (Intel Xeon Gold 6142, 2.6 GHz, 16-core chip), so the total time spent in one subset was about 10 hours using 320 CPU cores in an Intel processor-based HPC cluster. In addition, it spent about 3 hours to extract feature via deep neural network using 8 Tesla V100 GPUs of the NVIDIA DGX-1 system. In contrast to about 160 hours cost in traditional classification methods, this approach cost a little more than 200 hours using 8 Tesla V100 GPUs and 320 CPU cores.

For the same 2-million image dataset, cryoDRGN took about 29 hours for one epoch of training on 8 Tesla V100 GPUs, which amounted to 725 hours for training over 25 epochs. 3DVA took about 18 hours for one epoch of optimization with one Tesla V100 GPU. These observations indicate that the benefit of single GPU-based speedup with processing smaller datasets is largely negated by the problems of overloaded GPU memory and I/O bandwidth of network file systems when dealing with very large datasets. All methods tested herein exhibit formidable difficulties in keeping the same speedup ratio as observed on small datasets when processing the 2-million datasets, and suffer from an exponentially increased computational cost, which is a commonly recognized issue in high-performance computing. Taken together, these observations suggest that the computational cost and efficiency of AlphaCryo4D are within the acceptable range considering the output of more high-resolution conformers yielded by the procedures. The benefits likely outweigh the tradeoff of computational costs, particularly for very large datasets that equally slow down any methods.

**Relation to other techniques proposed to analyze structural heterogeneity**

Our particle-shuffling and Bayesian bootstrap strategy is conceptually related to the stochastic bootstrap method previously proposed for estimating 3D variance map[69] or performing 3D PCA[65] to detect conformational variability. One essential difference lies in that our bootstrap requires that only particles of similar or identical conformations via likelihood-based similarity estimation are grouped together instead of being combined or resampled in a stochastic manner[60,70]. Another difference is that the goal of $M$-fold particle shuffling is to improve the quality of reconstituted energy landscape and to enable energy-based particle voting for intrinsic cross-validation of 3D classification.

The application of multivariate statistical analysis directly on 3D density maps have been demonstrated to be useful for studying conformational changes[65,71,72]. As another step forward, the combination of 3D PCA[65,71] and energy landscape computation[40,51,52] has been shown to reasonably effective in capturing the overall conformational landscape. The potential benefit of using manifold embedding, such as diffusion map, instead of PCA, for reconstituting energy landscapes has also been demonstrated at low resolution, with presumption that the conformational changes can be discerned through a narrow angular aperture[5,53,54]. The presumption restricts its potential applications to complicated conformational dynamics.

The recently proposed 3DVA algorithm[21] implementing probabilistic PCA-based 3D classification in cryoSPARC, deep generative model-based cryoDRGN[22] and Laplacian spectral volumes method[73] were designed to tackle the problem of recovering continuous conformational changes without reconstituting energy landscapes. cryoDRGN uses a variational autoencoder (VAE) to map raw particle images to a feature space. The Laplacian spectral volume method is based on graph Laplacian representations of conformational heterogeneity and may also be considered as a generalization of PCA[73]. This algorithm was only examined with synthetic datasets. In addition, methods of integrating traditional molecular dynamics (MD) simulations with cryo-EM data, with or without use of deep learning, have been recently proposed to extract protein dynamics information[74,75]. These methods cannot be used to recover structures and dynamics hidden in the particle sub-dataset not being accurately classified and not used in final cryo-EM reconstructions and are limited by the size of biomolecules that can afford MD simulations.

**Advantages and limitations**. AlphaCryo4D exhibits several advantages as well as certain limitations. First, it does not assume any prior knowledge regarding the conformational landscape, its continuity and topology, as well as chemical composition of the macromolecules imaged. In contrast to other methods like 3DVA and cryoDRGN that use one consensus model for initial alignment of particle poses by cryoSPARC, our approach allows multiple consensus models being used for optimizing the particle alignment prior to the machine-learning-based reconstitution of energy landscape and 3D classification. This can theoretically deal with more complicated conformational dynamics when no stable core structure is available to guide the consensus alignment over a single model. Second, the choices of using energy landscape to represent the conformational variation has theoretical roots in physical chemistry and protein dynamics, in which extensive research tools being built around the energy landscape and transition-state theory allows correlation with other complementary approaches, such as single-molecule florescence microscopy, NMR and molecular dynamics simulation. The energy landscape allows users to examine kinetic relationships between the adjacent conformers and to potentially discover new conformations. Third, AlphaCryo4D was designed to optimize the resolution or 3D classification accuracy as the primary objective, in ways without losing the statistical power of characterizing the full spectrum of conformational states hidden in the sample. It allows a maximal number of particles to be assessed and classified in an integrative procedure based on uniform, objective criteria, such as their reproducibility and robustness in deep manifold learning. Last, the energy-based particle-voting algorithm can potentially rescue certain particles that are prone to be misclassified if processed only once by deep manifold learning, thus boosting the efficiency of particle usage without necessarily sacrificing quality and homogeneity of selected particles.

 The limitation of AlphaCryo4D lies in that it requires generally larger datasets and more computational costs to fully exploit its advantages and potentials. We do not expect to achieve considerably better results for a small dataset. A small dataset may be not enough to achieve high resolution for reconstructing conformational continuum by any means due to lack of structural signals. In fact, it is believed that the data size must proportionally scale with and match the degree of conformational heterogeneity in order to visualize a full spectrum of conformational states at the atomic level. Moreover, its outcomes are dependent on the success of initial consensus alignment of all particles during the first step of particle shuffling and volume bootstrapping. Certain conformational dynamics or heterogeneity can interfere with image alignment, in which case the performance of AlpahCryo4D will be restricted by the alignment errors in during consensus refinement. This problem is a common challenge for all existing methods developed so far. Our tentative solution is to use multiple consensus models for improving image alignment. Those methods using only one consensus model in image alignment will suffer from higher alignment errors and thus greater loss in performance in those complicated circumstances[5,21,22]. Failure in obtaining accurate alignment parameters can lead to futile, erroneous estimation of conformational heterogeneity in the subsequent steps no matter which methods are used.

**Applications to experimental cryo-EM datasets**

Three experimental cryo-EM datasets were processed using AlphaCryo4D to examine its applicability in analyzing experimental cryo-EM data, as described in detail in the following.

(1) *Substrate-free proteasome dataset*. The dataset of the ATPγS-bound human 26S proteasome[14] contains 455,680 particles and was previously used to determine six coexisting conformational states of the substrate-free proteasome. The cryo-EM data were recorded with the super-resolution pixel size of 0.75 Å. The dataset is publicly available in the Electron Microscopy Public Image Archive (EMPIAR) database at EMDataResource.org (accession code EMPIAR-10090). The size of ATPγS-bound proteasome dataset was enhanced to 455,680 × 3 particles and was divided into 4 sub-datasets for volume bootstrapping. 160 volumes were bootstrapped to compute the energy landscape the substrate-free ATPγS-bound proteasome (Fig. 3a).

(2) *Substrate-engaged proteasome dataset*. The substrate-engaged human proteasome dataset[2] includes 3,254,352 RP-CP particles (combined with particle images from both the DC and SC proteasomes) in total, with the super-resolution counting mode pixel size of 0.685 Å and the undecimated box size of 600 × 600 pixels. Sample preparation, cryo-EM imaging and data collection condition as well as data pre-processing prior to 3D classification by AlphaCryo4D have been previously described in detail[2]. All particles were binned twice down to 300 × 300 pixels prior to data processing. To proceed with AlphaCryo4D, the whole dataset was randomly split into 32 sub-datasets, each with 100,000 particles. Particle shuffling and volume bootstrapping procedures were conducted on each sub-dataset, which enhanced the total data size to 3,254,352 × 3 particles and yielded 32 × 4 × 10 = 1280 volumes in total that were used to compute the energy landscape of the substrate-engaged proteasome (Fig. 3b, c). Particle-voting on this energy landscape detected previously missing states (such as $E_{D0.2}$, $E_{D0.3}$, $E_{D1.1}$ and $E_{D1.2}$, etc.) and yielded previously determined states with improved map quality, such as $E_{A2}$ and $E_{D2}$. Guided by this initial discovery of new conformational states, extensive focused 3D classifications in AlphaCryo4D were performed, as described in detail in the next section, to visualize novel ubiquitin-binding sites and hidden dynamics of proteasomal AAA-ATPase motor during single-nucleotide exchange.

(3) *Inflammasome dataset*. The dataset of human NLRP3-NEK7 complex includes 622,562 particles[26]. All particles were divided into 6 sub-datasets for particle shuffling, with the total data size enhanced to 622,562 × 3 images for volume bootstrapping. Given the small molecular size of the complex, the super-resolution pixel size of 0.42 Å and the box size of 240 pixels were employed in image analysis. 480 volumes were bootstrapped to compute the energy landscape of the NLRP3-NEK7 complex (Extended Data Fig. 6l). The NLRP3-NEK7 dataset exhibits certain degree of orientation preference. While AlphaCryo4D was able to retrieve 3 distinct NLRP3 conformations, the resulting map resolution is limited to ~4 Å. Although this may be a potential improvement over the previous analysis of this dataset by the conventional approach, the result suggests that AlphaCryo4D does not necessarily overcome the potential issues of the orientation preference.

**Focused 3D classification by AlphaCryo4D**

To apply AlphaCryo4D to process the experimental dataset of the substrate-engaged human 26S proteasome[2], a focused 3D classification protocol using AlphaCryo4D was implemented to enhance its capability in discovering transient states of extremely low particle populations and in solving highly dynamic components that may be too small to be analyzed by conventional methods[6-8,17]. To study the local features of ubiquitin binding and substrate interactions, 3D masks corresponding to the local densities of interest were applied to all bootstrapped volume data, and the subset of interest on the energy landscape was extracted to conduct a zoomed-in analysis. In this case, the energy landscape calculated via masked volumes is expected to reflect the local structural variations of interest (Extended Data Fig. 6).

To detect novel ubiquitin-binding sites in the proteasome, local masks were applied to focus on the features around the putative locations of ubiquitin chains. After particle shuffling for $M = 3$ folds, $732,666 \times 3$ particle images of the state $E_A$ and $E_B$ were utilized for computing the zoomed-in energy landscape with 280 volumes. These particles were first aligned in a consensus alignment without any mask or resolution limit to restrict alignment parameter calculations. On the focused energy landscape, only the region within the soft mask of the ubiquitin and its binding site was applied in the volume bootstrapping step, with the signals of the CP of the proteasome subtracted from each raw particle when dealing with the state $E_A$ and $E_B$. Accordingly, the particle diameter applied in the CP subtracted proteasome was shrunk to 274 Å, instead of 411 Å in the complete 26S proteasome. When the region around RPN2 masked for focused 3D classification to search for potential ubiquitin-like features, the total particle numbers were enhanced to $208,700 \times 3$ ($E_{A1}$), $240,475 \times 3$ ($E_{A2}$), $260,021 \times 3$ ($E_B$) and $129,492 \times 3$ ($E_{D2}$). The energy landscapes were computed with 80, 80, 80 and 40 masked volumes bootstrapped by upper resolution bounds of 15, 15, 15 and 20 Å, respectively. Several 3D classes with the particle numbers of 192,219, 2,539, 3,982 and 609 were obtained by the particle-voting procedure on the zoomed-in energy landscape of state $E_{A1}$, while the case of state $E_{A2}$ gave four classes containing 147,108, 5,842, 5,500 and 33,083 particles (Extended Data Fig. 6e, f). Likewise, the particle numbers of 3D classes on the zoomed-in energy landscape of state $E_B$ were 173,931, 5,117 and 9,754 (Extended Data Fig. 6g); and the zoomed-in energy landscape of state $E_{D2}$ resulted in four classes having 61,580, 6,192, 24,674 and 1,673 particles, respectively (Extended Data Fig. 6h).

To detect and improve RPN1-bound ubiquitin densities, a local mask around RPN1 was used, with 15 Å resolution limit applied in the volume-bootstrapping step. The zoomed-in energy landscape around state $E_{A1}$ computed from 120 volumes resulted in five classes with the particle numbers of 128,161, 7,754, 6,557, 6,826 and 3,937. Five 3D classes on the zoomed-in energy landscape of $E_{A2}$, also computed with 120 volumes, including 81,612, 27,483, 12,247, 26,029 and 12,777 particles, were analyzed (Extended Data Fig. 6b, c). Based on the subclass of $E_{A1}$ that contains 128,161 particles, we continued to refine the alignment locally to the component of RPN1. Then a 3D mask around RPN1 was applied to bootstrap 40 volumes with 20 Å resolution limit in the expectation step. After voting on the energy landscape, two classes with the particles number of 43,957 and 43,620 were generated, one of which exhibited a ubiquitin-like density on the T1 site of RPN1 and was designated state $E_{A1.1}$ (Extended Data Fig. 6d). In the case of the ubiquitin-binding site on the α5 subunit of the CP, five 3D classes with the particle numbers of 136,071, 19,351, 4,100, 131,570 and 4,932 were generated by particle voting on the zoomed-in energy landscape of $E_D$ computed with 120 volumes, which were bootstrapped from 326,409 × 3 particles via the mask containing this ubiquitin-binding site and no high-resolution restriction (Extended Data Fig. 6i). During these analyses, all masked volume-bootstrapping steps were performed for 50 iterations, whereas unmasked 3D classifications were performed for 30 iterations with the initial reference low-pass filtered to 60 Å. At the last step, 3D refinement and reconstruction of each class was done to finalize the high-resolution structure determination.

To exploit intermediate conformations during the substrate translocation initiation, a zoomed-in energy landscape was computed for the focused 3D classification procedure. First, the energy landscape of all 3,254,352 × 3 particles was computed and analyzed. Then the zoomed-in energy landscape of the states $E_C$ and $E_D$ containing 2,521,686 × 3 particle images were computed from 1,000 volumes that were bootstrapped without masking or resolution restriction (Extended Data Fig. 6j). For further zoomed-in processes, the particles around the translocation initiation states (from $E_C$ to $E_{D0}$) on this energy landscape were extracted by the distance-based 3D classification strategy. Subsequently, these particles were augmented to the number of 410,098 × 3 to generate 160 volumes for the computation of the zoomed-in energy landscape with the RP mask and no resolution restriction imposed in the volume bootstrapping, which led to the 3D classes $E_{C1}$, $E_{C2}$, and $E_{D0.1}$ with the particle number of 68,506, 77,545, and 25,755, respectively (Extended Data Fig. 6k). Moreover, state $E_{D0.3}$ came from two clusters with 86,415 and 83,107 particles merged together on the unmasked energy landscape, and the classes $E_{D0.2}$ and $E_{D1.1}$ including 105,081 and 51,808 particles, respectively, were also obtained after particle voting on this energy landscape. In the local areas of the state $E_{D1}$ on the energy landscape, two clusters with 80,841 and 94,273 particles were merged to reconstruct the high-resolution structure of state $E_{D1.2}$, while states $E_{D1.1}$ and $E_{D1.3}$ with 40,696 and 145,506 particles, respectively, were separated out from the clusters on this unmasked energy landscape. In these studies, the maximum number of iterations in the 3D classifications of volume bootstrapping was 30, with the initial reference low-pass filtered to 60 Å routinely.

Beyond resolving significantly more conformers from the same dataset, AlphaCryo4D also allows us to push the envelope of the achievable resolution of dynamic components and key metastable states due to its advantage in keeping more particles without sacrificing conformational homogeneity. To improve the resolution of state $E_{D2}$, a rigorous particle voting procedure was performed in each cluster on the unmasked energy landscape (Fig. 3b, c), which improved the overall quality of particle dataset while classifying significantly more particles to this $E_{D2}$ conformational class. Next, we conducted a focused 3D classification by using AlphaCryo4D to calculate the zoomed-in energy landscape of the entire $E_{D2}$ dataset to detect any potential conformational changes within this class. Five clusters near the energy wells of state $E_{D2}$, whose particle numbers were 154,661, 243,799, 198,221, 179,257 and 106,047, were investigated by separate high-resolution 3D refinement and reconstruction. The resulting refined density maps at 3.0-3.5 Å were found to be all in an identical conformation as previously reported[2]. These five clusters were then merged together to refine a final 3D density map of state $E_{D2}$. Thus, AlphaCryo4D allowed us to obtain a new $E_{D2}$ dataset that appears to retain high conformational homogeneity while including three-fold more particles, which improves its 3D reconstruction from the previously published resolution at 3.2 Å to 2.5 Å, measured by gold-standard Fourier shell correlation (FSC) at 0.143-cutoff. By contrast, in our previous work[2], we had attempted to include more $E_{D2}$-compatible 3D classes from 3D maximum-likelihood classification in RELION, which resulted in lower resolution at 3.3 Å by doubling the $E_{D2}$ dataset, indicating that including more particles by conventional methods gave rise to higher heterogeneity in the class presumably due to increased misclassification.

**Analysis of doubly and singly capped proteasomes**

Previous cryo-EM analyses of the proteasomal states were nearly all focused on the RP-CP subcomplex, by converting double-cap (DC) proteasome particles to pseudo single-cap (SC) proteasome, which was always mixed with true SC particle images, in order to improve the resolution with conventional methods[2,13-15,76,77]. As a result, the cryo-EM density of the CP gate in the other side distal to the RP in the RP-CP subcomplex has been an average of closed and open states with unknown stoichiometry. This has made it impossible to appreciate the difference between the SC and DC proteasomes in the previous studies. To remove this ambiguity, we reconstructed 8 density maps of the true SC proteasome and 36 density maps of the true DC proteasome including all possible combination of the 8 major RP-CP states. To reconstruct the density maps of the DC proteasome, pseudo-SC particles corresponding to 8 different RP states were voted out from the energy landscape, with the sub-states belonging to the same RP state combined together. Then the particles with one RP-CP were extracted from the particle stacks to refine the RP and CP density maps with RP and CP contour masks applied in the local search step in RELION, respectively. For each reconstruction of the DC proteasome, two RP density maps on the opposite sides of the CP were expanded from 600 × 600 × 600 pixels to 800 × 800 × 800 pixels using RELION and aligned by the density of CP in Chimera before being merged together.

Counting the particle number of the 36 states of the DC proteasome, we could derive the distribution of the DC states, $p_{ij}$, and that of RP $p_i$ only, where $i$ and $j$ denote the states of two RPs in the same DC proteasome. Based on the value of $p_{ij}$ and total particle number, the experimental state distribution of the DC proteasome was plotted for quantitative analysis in two dimensions with respect to the states of two RPs in the same DC proteasome (Fig. 3e). If the states of two RPs of the same DC proteasome were independent of each other, the predicted state distribution $\tilde{p}_{ij}$ of the DC proteasome can be calculated as:

$$\tilde{p}_{ij} = p_i \cdot p_j$$

By using the total particle number and experimentally measured $p_i$, the 2D state distribution of the DC proteasome with two uncoupled RPs was calculated for further comparison to investigate the conformational entanglement effect of the two RP states (Fig. 3f-h).

**Atomic model building and refinement**

To build the initial atomic model of the newly discovered states, we use previously published proteasome structures as a starting model and then manually improved in Coot[78]. For the conformational states at resolutions lower than 5 Å, pseudo-atomic modelling was conducted in the following steps. First, each subunit was fitted as a rigid body, often in UCSF Chimera[79] as well as in Coot[78]. Then, the local structural domains and secondary structure elements were fitted as a rigid body, with linking loops flexibly fitted and modelled. After the mainchain structures were well fitted, no further fitting of the sidechain rotamers were pursued due to insufficient resolution except for energy minimization to remove unrealistic side-chain clashes. For the conformational states at resolutions higher than 5 Å, the mainchain traces were first fit in Coot.

Then, extensive fitting, adjustment and optimization of sidechain rotamers were conducted through local real-space refinement and manual rectification in Coot to the degree commensurate to the map resolution and quality. Atomic model refinement was conducted in Phenix[80] with its real-space refinement program. We used both simulated annealing and global minimization with NCS, rotamer and Ramachandran constraints. Partial rebuilding, model correction and density-fitting improvement in Coot[78] were iterated after each round of atomic model refinement in Phenix[80]. The improved atomic models were then refined again in Phenix, followed by rebuilding in Coot[78]. The refinement and rebuilding cycle were often repeated for three rounds or until the model quality reached expectation (Extended Data Table 2). All figures of structures were plotted in Chimera[79], PyMOL[81], or ChimeraX[82]. Local resolutions of cryo-EM density maps were evaluated using ResMap[83] or Bsoft Blocres program[84]. Structural alignment and comparison were performed in both PyMOL and Chimera. Electrostatic surfaces were calculated by APBS plugin[85] in PyMOL.

Given a large range of variations of local resolution in an expanded number of proteasome states and substates, as well as reconstructions with different stoichiometric ratios of RP versus CP, additional caveats and cautions were practiced in order to avoid misinterpretation and overall-fitting of atomic and pseudo-atomic models. First, when fitting the ubiquitin structure to the low-resolution local density, we consider both the existing homologous structural models as a modelling reference. Specifically, the fitting of diubiquitin to the density on RPN2 in states $E_{A2.1}$, $E_{A2.2}$, and $E_{B.2}$, we took the NMR structure of the yeast diubiquitin-bound Rpn1 T1 site as a reference, because of the high structural homology between RPN1 and RPN2. For the fitting of RPN10-bound density, we also took into account the electrostatic complementarity (Extended Data Fig. 10). Second, even within an overall high-resolution cryo-EM map, it often presents certain local densities at a lower local resolution due to poor occupancy of ligands or flexibility of less well-folded segments. For example, for the nucleotide in RPT5 in states $E_{D0.1}$ to $E_{D1.3}$, the high-resolution atomic model of ADP refined from other states like $E_{C1}$ were used to fit these lower-local-resolution densities as a rigid body. However, the B-factors of the fitting atoms of ADP in RPT5 were reported to be more than twice higher than those of other nucleotides, indicating its partial occupancy or unstable binding. Thus, ADP fitting in RPT5 should be regarded tentative for the most or for the purpose of evaluation of nucleotide states rather than a reliable high-resolution atomic modeling. Similarly, the atomic models of these locally low-resolution features were only further adjusted with strict stereochemical constraints when the map resolution and features permit. Third, for the CP gate at the other side opposite to the RP in the RP-CP subcomplex reconstructed from a mixture of true and pseudo-SC proteasomes (converted from the DC proteasomes), their local densities in states $E_{D0.1}$ to $E_{D2}$ appears to be an average of the open and closed states, and thus are weaker in amplitude than the rest of the CP. In the atomic model building of these reconstructions, we chose to model them with the closed CP gate state as long as their density resolution allows atomic modeling. This issue was completely solved when we separated the 3D classes of the SC and DC proteasomes and reconstructed their states respectively, where the atomic modelling of any CP gate no longer suffers from such ambiguities.

**Quantum mechanical calculation**

First-principles calculations on the electronic structures of the substrate-proteolytic site complex were conducted as described. In brevity, local coordinates of the substrate-bound CP near the catalytic site at residue Thr1 were taken out from the experimentally fitted atomic models of CP and the substrate, in both $E_A$ and $E_{D2}$ conformations. The proteolytic site is the Thr1 residue at the β2 subunit but extended along the chain for another two residues to make the system less finite. Hydrogen atoms are added to complete the residues, and to saturate the system boundaries. The structural relaxations and electronic structure calculation were carried out using the density functional theory (DFT) method with norm-conserving pseudopotentials as implemented in the SIESTA code[86]. To set up the initial conditions of the finite system, the hydrogen atom in the hydroxy group of Thr1 is manually relocated to the nearby nitrogen atom in the same residue, corresponding to the initial proton transfer in the catalytic mechanism. All hydrogen atoms were subsequently optimized without symmetric restrictions using the conjugate gradient algorithm and a 0.04 eV/Å maximum force convergence criterion while keeping the rest of the system fixed. The generalized gradient approximation (GGA) exchange-correlation density functional PBE[43] was employed together with a double-zeta plus polarization basis set, and a mesh cutoff of 200 Ry (corresponds to 0.23 Å smallest grid size). The charge density and charge density difference contour maps were plotted with the Siesta Utility programs denchar and Python, respectively.

**Data availability**

The three-dimensional cryo-EM density maps of all new states are deposited into the Electron Microscopy Data Bank (EMDB) (www.emdatasource.org) under accession numbers to be provided upon formal publication of this manuscript). The coordinates are deposited in the Protein Data Bank (PDB) (www.wwpdb.org) (with accession numbers to be provided upon formal publication of this manuscript). Raw data are deposited into the Electron Microscopy Pilot Image Archive (www.ebi.ac.uk/pdbe/emdb/empiar) (with accession numbers to be provided upon formal publication of this manuscript). Source code of AlphaCryo4D is freely available for download (http://ipccsb.dfci.harvard.edu/alphacryo4d/).

**Acknowledgments.** The authors thank Q. Ouyang, A. Goldberg, D. Finley, S. Elsasser, M. Kirschner, Y. Lu, and for constructive discussions. This work was funded in part by the Natural Science Foundation of Beijing Municipality (grant No. Z180016/Z18J008), the National Natural Science Foundation of China (grant No. 11774012) and a gift academic grant from Intel Corporation. The cryo-EM data were collected at the Cryo-EM Core Facility Platform and Laboratory of Electron Microscopy at Peking University. The data processing was performed in the High-Performance Computing Platform at Peking University.

**Author contributions**. Y.M. and Z.W. conceived this study, devised the methodology and wrote the paper. Z.W. developed the prototype source code of the AlphaCryo4D system and conducted numerical studies of the system using the synthetic datasets. S.Z., Y.D., Z.W. and W.L.W. contributed to the experimental cryo-EM data acquisition. Z.W. and S.Z. analyzed the experimental cryo-EM datasets and refined the density maps. W.L.W. conducted the quantum mechanical simulation and contributed to the improvement of the AlphaCryo4D source code. Y-P.M contributed to early studies in algorithm design and testing. Y.M. supervised this study, verified the density maps, built and refined the atomic models, interpreted the data and drafted the manuscript with inputs from all authors.

**Competing interests**. The authors declare no competing financial interests.

**Correspondence and requests for materials** should be addressed to Y.M.

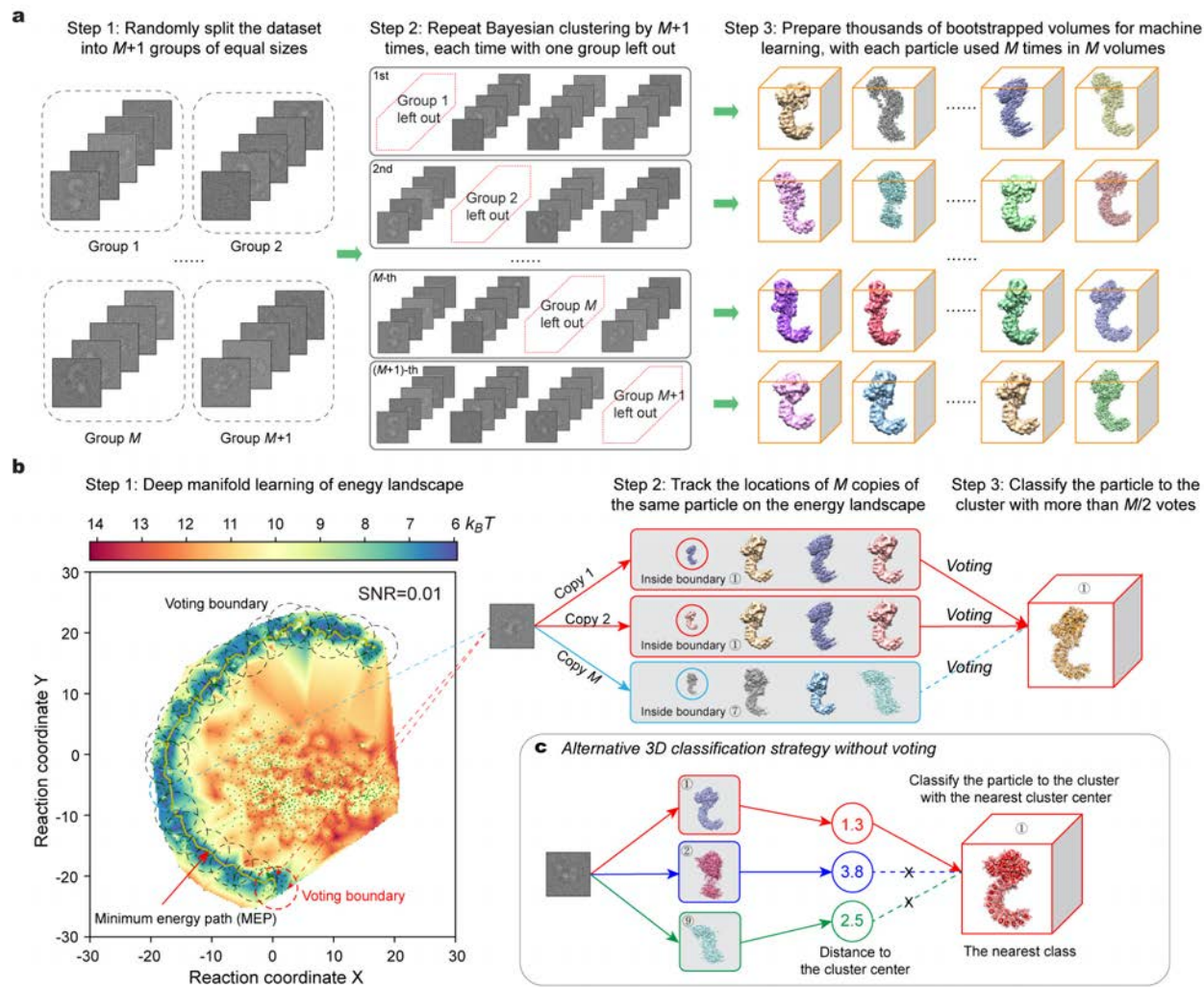

**Extended Data Fig. 1. Detailed algorithmic design of particle shuffling and voting in AlphaCryo4D. a**, Schematic showing the method of *M*-fold particle shuffling and bootstrapping of 3D volumes. In the step 1, all particles are split randomly into *M* + 1 groups equally. Then the step 2 carries out the Bayesian clustering for reconstructions of 3D density maps within each of the *M* + 1 particle sets that are shuffled via the leave-one-group-out approach. After these two steps, thousands of 3D volumes are generated for the subsequent 3D deep learning, with each particle contributing to *M* volumes. **b**, Schematic showing the algorithmic concept of energy-based particle voting for 3D classification. The left panel shows the energy landscape obtained by deep manifold learning. After clustering along the minimum energy path, all *M* locations of each particle on the energy landscape can be tracked to cast *M* votes. A vote of the particle is only counted for the cluster when it is mapped within the voting boundary of the cluster, as indicated by the circles marked on the energy landscape. Eventually, this particle is classified to the 3D cluster with over *M*/2 votes from this particle. **c**, Alternative distance-based 3D classification method as a control in the analysis of algorithmic performance of particle voting. Instead of particle voting, each particle is directly classified to the cluster with the nearest clustering center among the M volume data points in the distance-based 3D classification strategy.

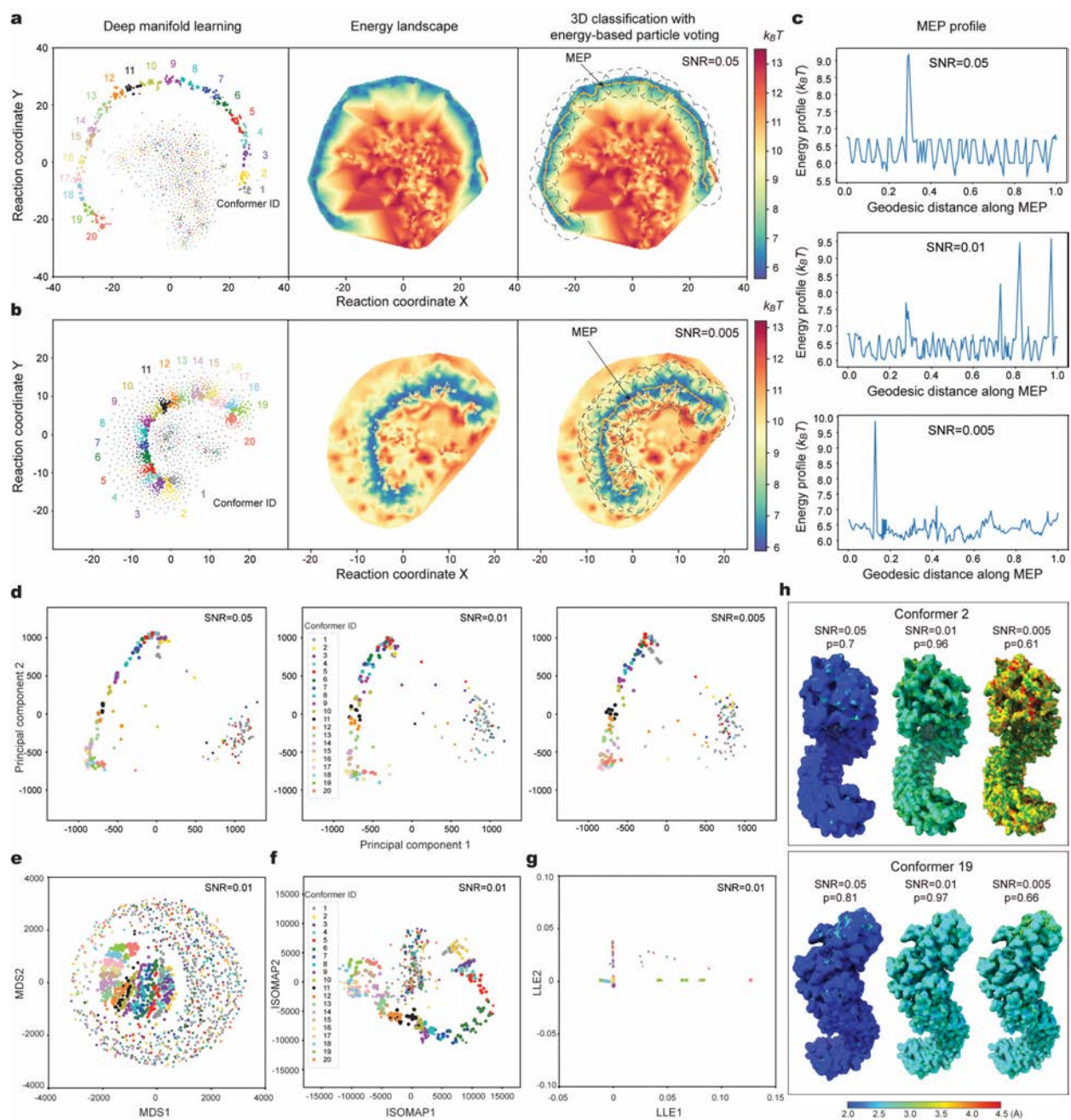

**Extended Data Fig. 2. Blind assessments of AlphaCryo4D and its comparison with the 3D PCA method using the simulated heterogeneous NLRP3 datasets of different SNRs. a** and **b**, Reconstruction of the energy landscape of the simulated NLRP3 datasets at SNRs of 0.05 (**a**) and 0.005 (**b**) by the t-SNE algorithm using the bootstrapped volumes and their corresponding feature maps. Colors in the left panels indicate the ground truth of 3D volume data points. **c**, Free energy profiles along the MEP calculated by the string method in the 2D energy landscapes of the simulated NLRP3 datasets at SNRs of 0.05 (left), 0.01 (middle) and 0.005 (right). **d**, Linear

dimensionality reduction of bootstrapped 3D volumes from the simulated datasets at three distinct SNRs by the 3D PCA method. **e-g**, Dimensionality reduction of bootstrapped 3D volumes from the simulated dataset at SNR of 0.01 by the multidimensional scaling (MDS)[66] (panel **e**), isometric mapping (Isomap)[67] (panel **f**) and locally linear embedding (LLE)[68] algorithms (panel **g**). Colors of data points indicate the ground truth of their corresponding 3D volumes. **h**, Comparison of local resolution assessment of AlphaCryo4D-classified NLRP3 reconstructions of conformers 2 (upper inset) and 19 (lower inset) from the simulated datasets of three distinct SNRs. The local resolutions were computed by ResMap[83]. Conformers 2 and 19 have been missed 10 and 8 times in 18 control tests using several other methods (see Extended Data Fig. 3b-g), which are the most and second-most frequently missed conformers, respectively. The 3D classification precision (p) is labelled above each density map. The color bar of local resolution is shown in the lower insert.

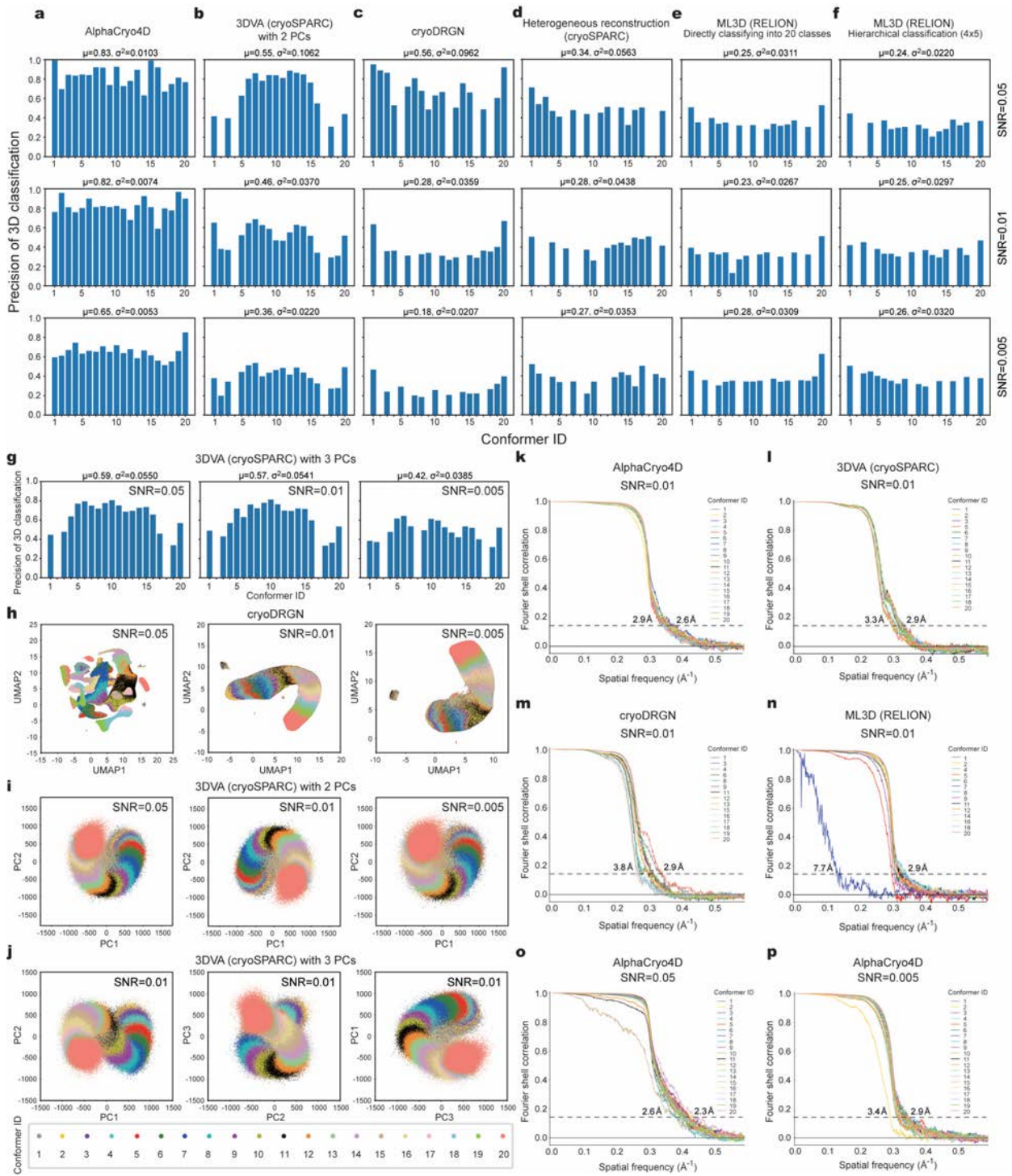

**Extended Data Fig. 3. Performance comparison of AlphaCryo4D with alternative methods using the simulated heterogeneous NLRP3 datasets of different SNRs. a-g**, 3D classification precision of the simulated datasets by AlphaCryo4D (**a**), 3DVA with two principal components (PCs) in cryoSPARC[21] (**b**), cryoDRGN[22] (**c**), heterogeneous reconstruction in cryoSPARC[8] (**d**)

and maximum-likelihood classification in RELION[6] (**e, f**). The results of SNRs of 0.05 (the first row), 0.01 (the second row) and 0.005 (the third row) are shown on three rows for side-by-side comparison. On the top of each panel, the symbols of $\mu$ and $\sigma^2$ denote the mean and variance of precision, respectively, with the values of missing classes treated as zeros. In the maximum-likelihood classification of RELION, both direct and hierarchical strategies are compared in the study. **g**, 3D classification precision of the simulated datasets by 3DVA using three PCs. **h-j**, Visualization of 3D classification by cryoDRGN in autoencoder-learned feature space (**h**), and by 3DVA with 2 PCs (**i**) and 3 PCs (**j**). **k-n**, The gold-standard FSC plots of the 20 maps resulting from 3D classification by AlphaCryo4D (**k**), of the 18 maps resulting from the 3D classification by 3DVA in CryoSPARC (**l**), of the 15 maps resulting from cryoDRGN (**m**) and of the 14 maps resulting from the maximum-likelihood 3D classification in RELION (**n**) on the simulated data of 0.01 SNR. They correspond to the precision results presented in the second row of panels (**a**), (**b**), (**c**) and (**e**), respectively. **o-p**, Additional gold-standard FSC plots of the refined density maps resulting from AlphaCryo4D on the simulated datasets of SNRs of 0.05 and 0.005.

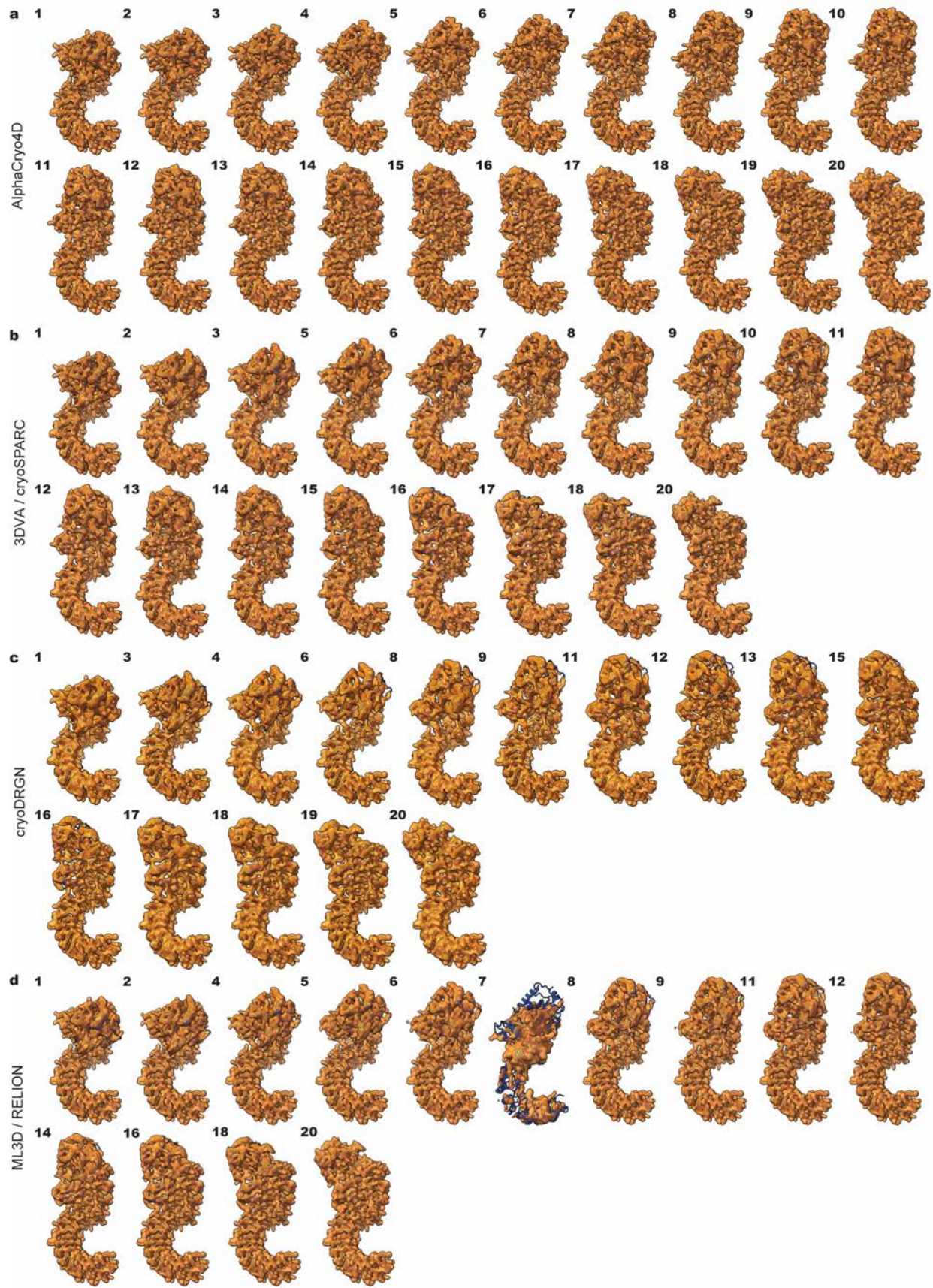
66

**Extended Data Fig. 4. Map assessments of AlphaCryo4D in reconstructions of conformational continuum in comparison with conventional methods on the simulated data of 0.01 SNR. a**, The 20 maps of distinct NLRP3 conformers resulting from the 3D classification by AlphaCryo4D. **b**, The 18 maps of NLRP3 resulting from 3DVA[21] in cryoSPARC. **c**, The 15 maps of NLRP3 resulting from the 3D classification by cryoDRGN[22]. **d**, The 14 maps of NLRP3 resulting from the 3D classification by ML3D in RELION[6]. All maps are shown in transparent surface representations superimposed with their corresponding atomic models of the ground truth in cartoon representations, which are fitted to the maps as rigid bodies without further atomic modelling. The conformer ID numbers are marked on the upper left of each map panel. The results shown in panels (**a**), (**b**), (**c**) and (**d**) correspond to the FSC results shown in panels (**k**), (**l**), (**m**) and (**n**) of Extended Data Fig. 3, respectively.

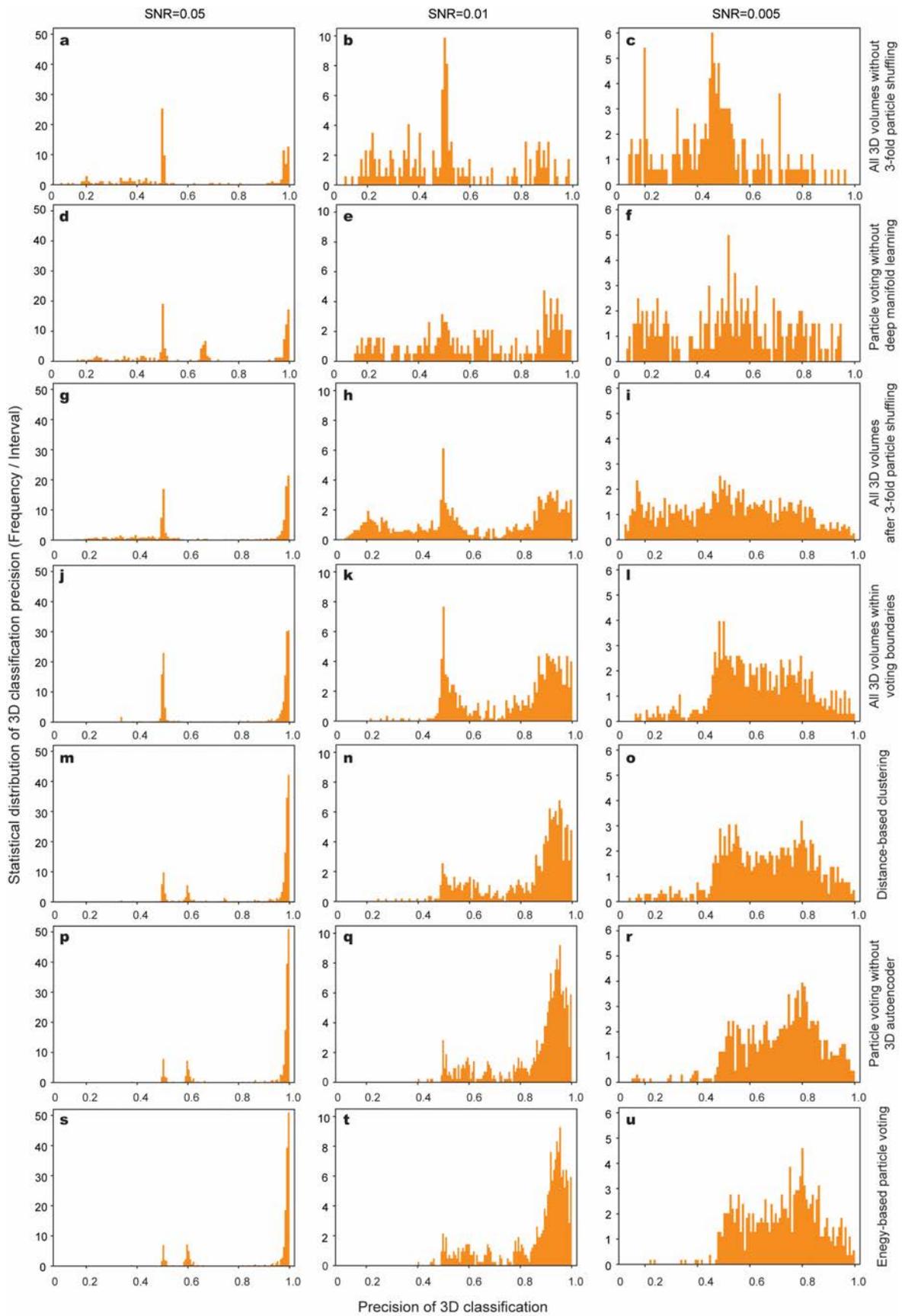

**Extended Data Fig. 5. Mechanistic characterizations of the improvement of 3D classification accuracy by AlphaCryo4D using the simulated NLRP3 datasets of three typical SNRs.** Left, middle and right vertical columns show the precision analysis on the simulated datasets with SNRs of 0.05, 0.01 and 0.005, respectively. In total, there are 18 conditional controls under six different algorithmic design variations being analyzed in panel (**a**)-(**r**). **a-c**, Statistical distribution of 3D classification precision in bootstrapped 3D volumes without particle shuffling. **d-f**, Statistical distribution of 3D classification precision by implementing particle voting directly on the bootstrapped 3D volumes without using deep manifold learning. The procedure of 3-fold particle shuffling and volume bootstrapping is identical to AlphaCryo4D. **g-i**, Distribution of 3D classification precision in bootstrapped 3D volumes after 3-fold particle shuffling in the intermediate step of AlphaCryo4D. **j-l**, Distribution of 3D classification precision in bootstrapped 3D volumes screened by voting boundary on the energy landscape in the intermediate step of AlphaCryo4D. **m-o**, Distribution of 3D classification precision after distance-based 3D clustering in the absence of energy-based particle voting with all prior steps identical to AlphaCryo4D. **p-r**, Distribution of 3D classification precision after particle voting by a modified AlphaCryo4D variation only without using deep residual autoencoder in the manifold embedding step. **s-u**, Distribution of 3D classification precision after energy-based particle voting via a complete AlphaCryo4D procedure.

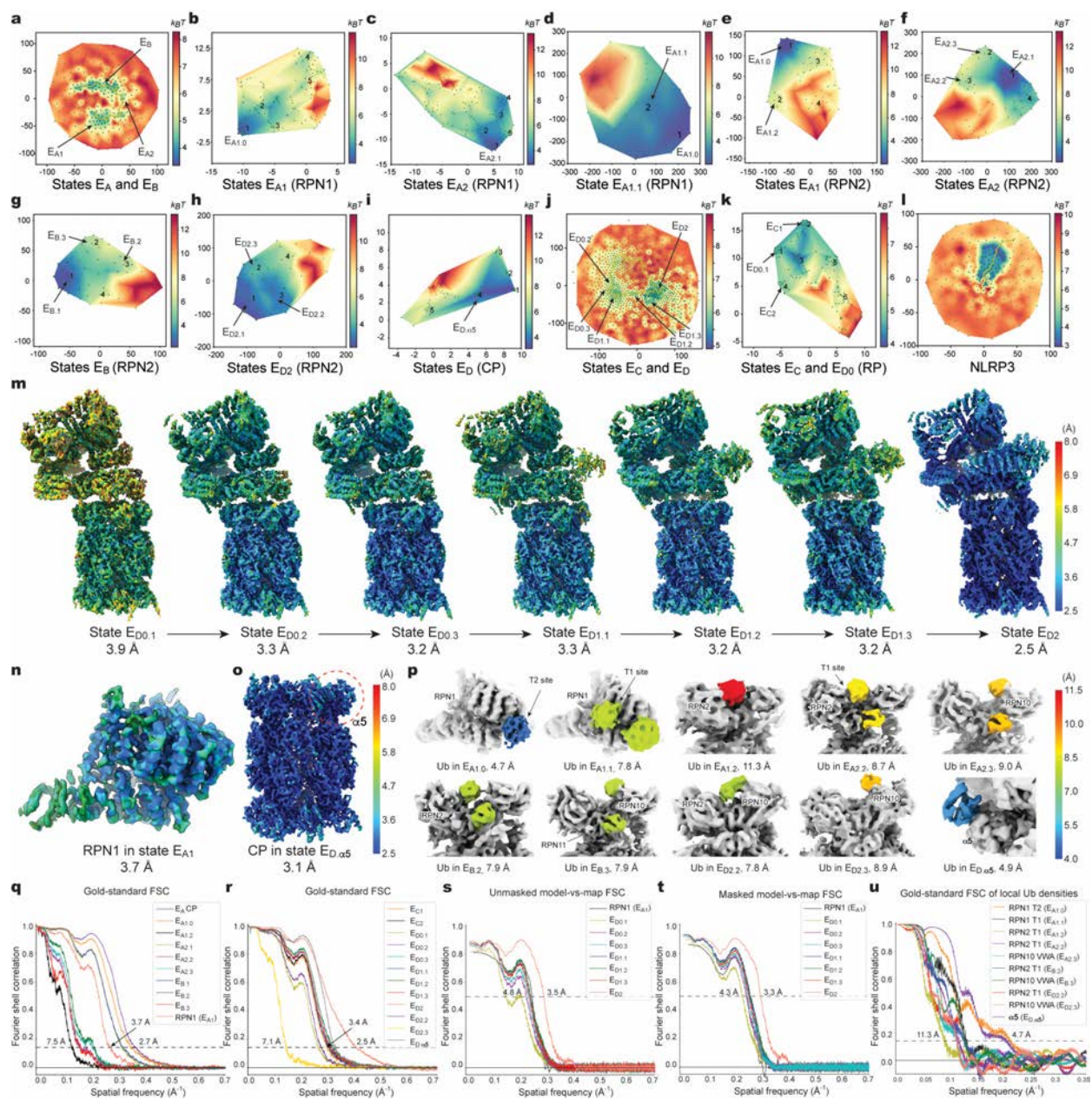

**Extended Data Fig. 6. Energy landscapes and cryo-EM reconstructions of novel proteasomal states determined by AlphaCryo4D. a-j,** The zoomed-in energy landscape used for focused 3D classifications of AlphaCryo4D on the experimental cryo-EM datasets for finding new states of the substrate-bound proteasome. The zoomed-in local energy landscapes of the substrate-bound human proteasome by AlphaCryo4D are shown, respectively, for state $E_A$ and $E_B$ in (**a**), for states $E_{A1}$ and $E_{A2}$ with 3D mask focusing on RPN1 in (**b**) and (**c**) to visualize the RPN1 T2 site, in (**d**) to visualize the RPN1 T1 site, for states $E_{A1}$, $E_{A2}$, $E_B$ and $E_{D2}$ with 3D masking around RPN2 in (**e**), (**f**), (**g**) and (**h**) to visualize the RPN2 T1 site, for combined states

of $E_D$ with masking around CP (**i**) to visualize the ubiquitin-binding sites on the α5-subunit, for states $E_C$ and $E_D$ in (**j**) and for combined states of $E_C$ and $E_{D0}$ with 3D mask focusing on the RP (**k**) to visualize the ATPase motor intermediates during translocation initiation. **l**, The energy landscape computed from the experimental cryo-EM dataset of the NLRP3-NEK7 complex. **m**, Local resolution measurement of the seven sequential intermediate states during translocation initiation measured by ResMap[83]. **n**, The local resolution measurement of the improved density map of RPN1 in state $E_{A1}$ by ResMap. **o**, The local resolution measurement of the CP in state $E_{D.α5}$ by ResMap. **p**, Putative ubiquitin (Ub) densities in ten distinct states colored by the local gold-standard FSC resolutions of the Ub densities measured in panel (**u**), with the rest of densities shown in light grey. The color bar for local resolutions is shown on the right in panels in **m-p**, where panels (**n**) and (**o**) share the same resolution color bar. **q** and **r**, The gold-standard FSC plots of the 20 conformers of the RP-CP subcomplex. **s** and **t**, The FSC curves calculated between the experimental cryo-EM maps and their corresponding atomic models for newly discovered states or improved states without (**s**) and with (**t**) masking in the calculations. **u**, The gold-standard FSC curves calculated between the two half maps with local Ub masks in ten distinct states.

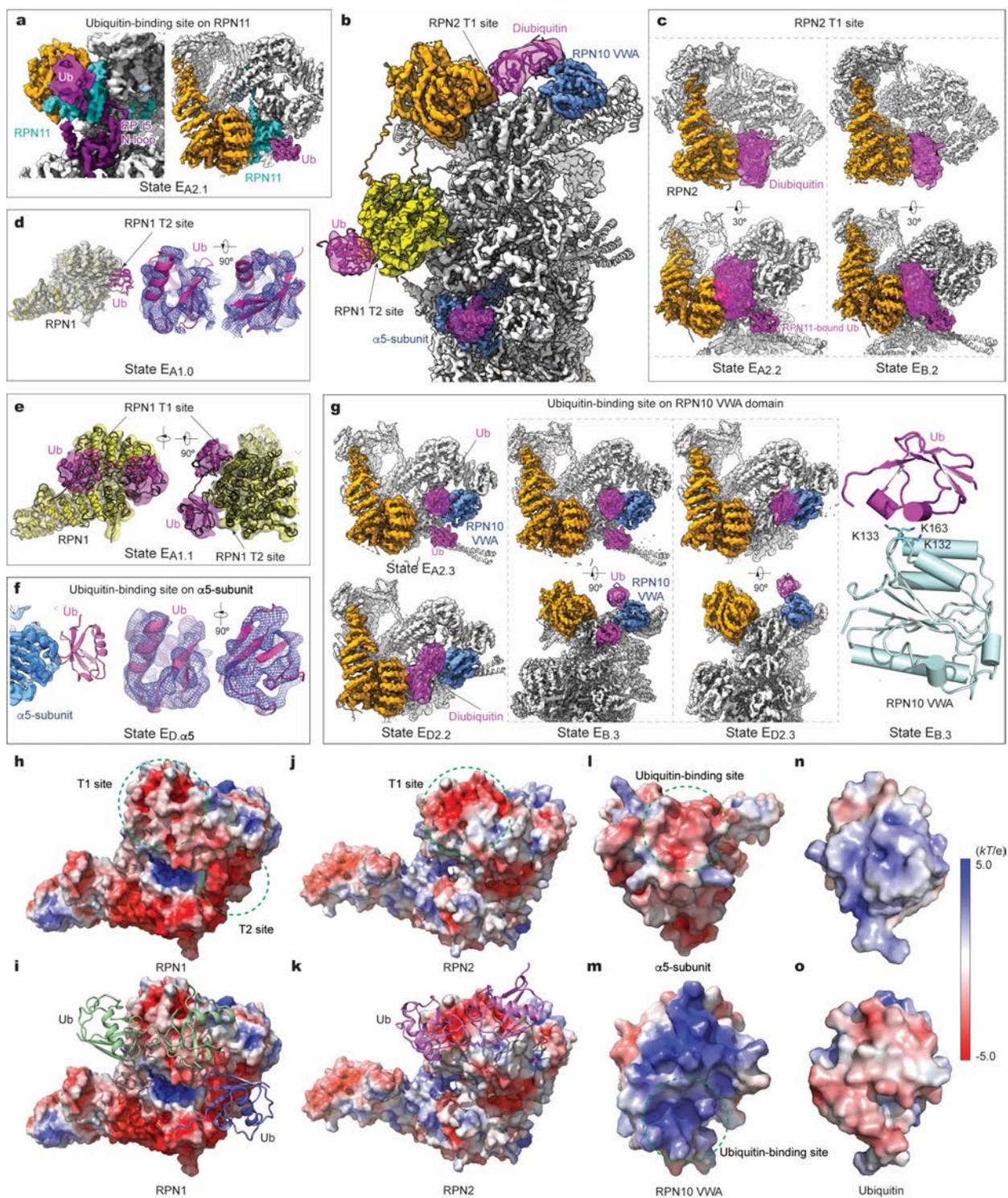

**Extended Data Fig. 7. Structural analysis of ubiquitin-proteasome interactions. a**, Cryo-EM density of state $E_{A2.1}$ shows the improved resolution localized around RPN11-bound ubiquitin and the RPT5 N-loop as compared to previously determined state $E_{A2}$. Ub, ubiquitin. **b**, High-resolution cryo-EM density of state $E_{D2}$ superimposed with Ub densities in state $E_{D2.2}$ highlighted

as magenta, showing that a diubiquitin chain in contact with both the RPN2 T1 site and the RPN10 VWA domain. The atomic model of state $E_{D2.2}$ in cartoon representation is superimposed with the density map. **c**, Closeup view of the RPN2-bound diubiquitin density in states $E_{A2.2}$ (left column) and $E_{B.2}$ (right column) in two different perspectives, with the lower row showing the coexistence of the RPN11-bound Ub and the RPN2-bound diubiquitin. **d**, The refined 3.7-Å RPN1 density map from state $E_{A1.0}$ in transparent surface representation superimposed with its atomic model with Ub bound near the RPN1 T2 site. The right insets show the local cryo-EM density of the RPN1-bound Ub in state $E_{A1.0}$ superimposed with its atomic model in cartoon representation. **e**, Cryo-EM density of state $E_{A1.1}$ in transparent surface representation superimposed with its pseudo-atomic model of cartoon representation, with one Ub moiety bound to the RPN1 T1 site and another Ub near the T2 site. The two Ub moieties appear to be compatible with a Lys63-linked diubiquitin model in an extended conformation, which is supported by the low-level density connecting the two Ub densities. **f**, Closeup view of the α5-subunit bound with Ub in the refined 3.1-Å density map of state $E_{D.α5}$. The cryo-EM density of the α5-subunit is shown in transparent surface representation superimposed with its atomic model. The right insets show the local cryo-EM density of the α5-bound Ub of state $E_{D.α5}$ in blue mesh representation low-pass filtered at 4.9 Å and superimposed with its atomic model in cartoon representation. **g**, Closeup views of the RPN10-bound Ub densities in states $E_{A2.3}$, $E_{B.3}$, and $E_{D2.3}$. Right panel shows the pseudo-atomic model of Ub interaction with the RPN10 VWA domain. The residues of RPN10 in contact with ubiquitin are labeled and shown as stick representation. Note that the Ub-binding mode of the RPN10 VWA domain here is different from that previously observed in the crystal structure of ubiquitylated yeast Rpn10 that also suggested Rpn10 VWA domain as a ubiquitin-binding domain[29]. In the yeast Rpn10, Ub binds to the Rpn10 surface area interfacing Rpn9 and sterically occluded Rpn10 incorporation into the proteasome[29]. **h-m**, The electrostatic surfaces of the RPN1, RPN2, RPN10 and α5 subunits show the charge complementarity of the RPN1 T1 and T2 (panels **h**, **i**), RPN2 T1 (panels **j**, **k**) and α5-binding (panel **l**) sites of Ub interactions are all acidic. By contrast, the RPN10 VWA site is basic (panel **m**), suggesting that it may bind the acidic side of Ub. **m** and **n**, The electrostatic surface of Ub shows one side that is basic (panel **n**) and the other side acidic (panel **o**).

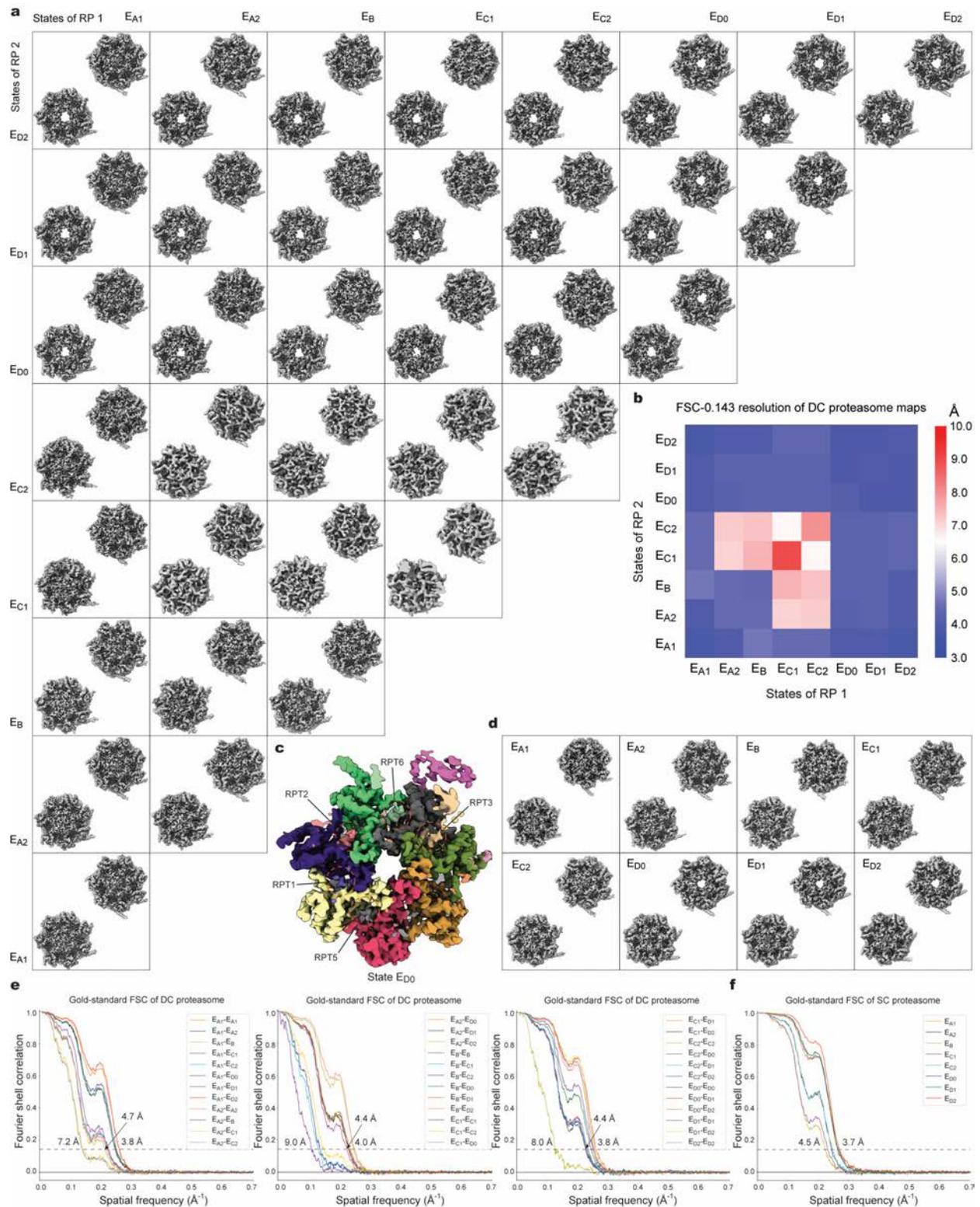

**Extended Data Fig. 8. Cryo-EM analysis of the doubly capped (DC) and singly capped (SC) proteasomes. a,** The CP gate states of the 36 density maps of distinct conformational states of the DC proteasome. All density maps shown are low-pass filtered to their measured resolutions

by the gold-standard FSC-0.143 cutoff without amplitude correction of B-factors. **b**, The FSC-0.143 resolution matrix of the 36 DC proteasomal states. **c**, The RP-CP interface of state $E_{D0}$, showing five C-terminal tails inserted into the inter-subunit surface pockets of α-ring. **d**, The CP gate states of the 8 density maps of distinct conformational states of the SC proteasome, with the upper right image in each panel corresponding to the RP-proximal side of the CP gate, and the lower left image to the RP-distal side of the CP gate. **e**, The gold-standard FSC plots of the 36 conformers of the DC proteasome. **f**, The gold-standard FSC plots of the 8 conformers of the SC proteasome.

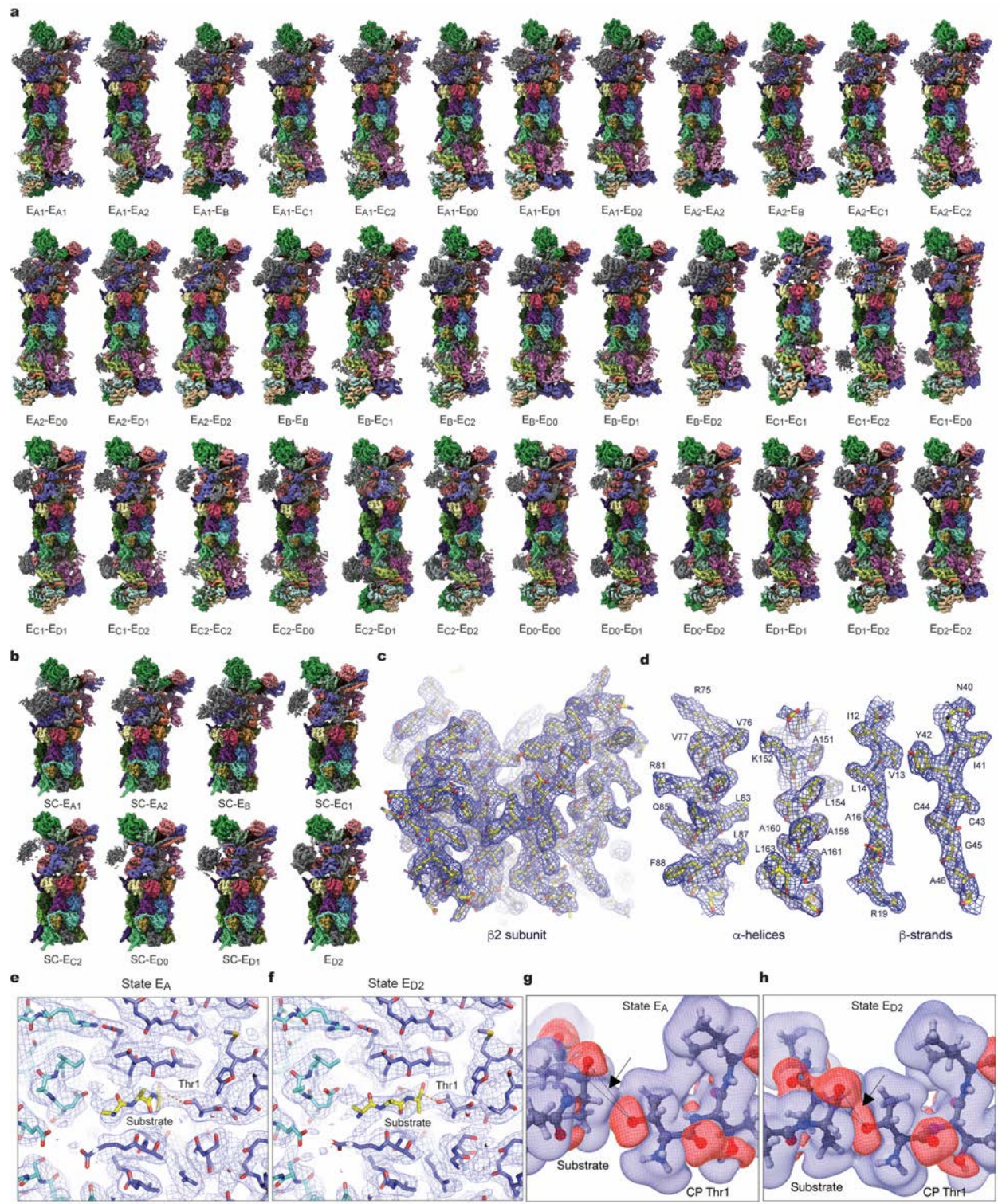

**Extended Data Fig. 9. Cryo-EM reconstructions of the SC and DC proteasomes and atomic-level analysis of substrate interactions with the CP active site in distinct states. a**, The density maps of 36 distinct conformers of the DC proteasome classified by AlphaCryo4D. **b**,

The density maps of 8 distinct conformers of the SC proteasome. All maps are filtered to their respective gold-standard FSC-0.143 resolutions without amplitude correction of B-factors and are differentially colored by their subunits in ChimeraX[82]. **c**, Closeup view of the cryo-EM density of the β2-subunit in state $E_{D2}$ shown in blue mesh representation superimposed with the atomic model. **d**, Cryo-EM densities of the typical secondary structure elements in the β2-subunit of state $E_{D2}$ shown as blue mesh representation superimposed with their corresponding atomic models. The residues are labelled. The densities are shown at 4σ level and exhibit structural features consistent with 2.5-Å resolution. **e**, The atomic model of substrate-bound catalytic active site of the β2-subunit superimposed with the cryo-EM density map of state $E_A$ at 2.7 Å. **f**, The atomic model of substrate-bound catalytic active site of the β2-subunit superimposed with the cryo-EM density map of state $E_{D2}$ at 2.5 Å. **g** and **h,** 3D Charge density difference map showing the interactions between the substrate polypeptide and the residue Thr1 in the β2-subunit of the CP. The red color labels an iso-surface of charge increase while the blue color labels charge decrease, both at a level of 0.04 e/Å$^2$. The charge increase region at the Thr1-Oγ atom is elongated sideways, and the substrate position in the $E_{D2}$ state provides both a proximity and orientation alignment for the pair to interact. The dashed line triangle shows the plane in which the detailed charge difference contours are plotted as shown in Fig. 6c, d.

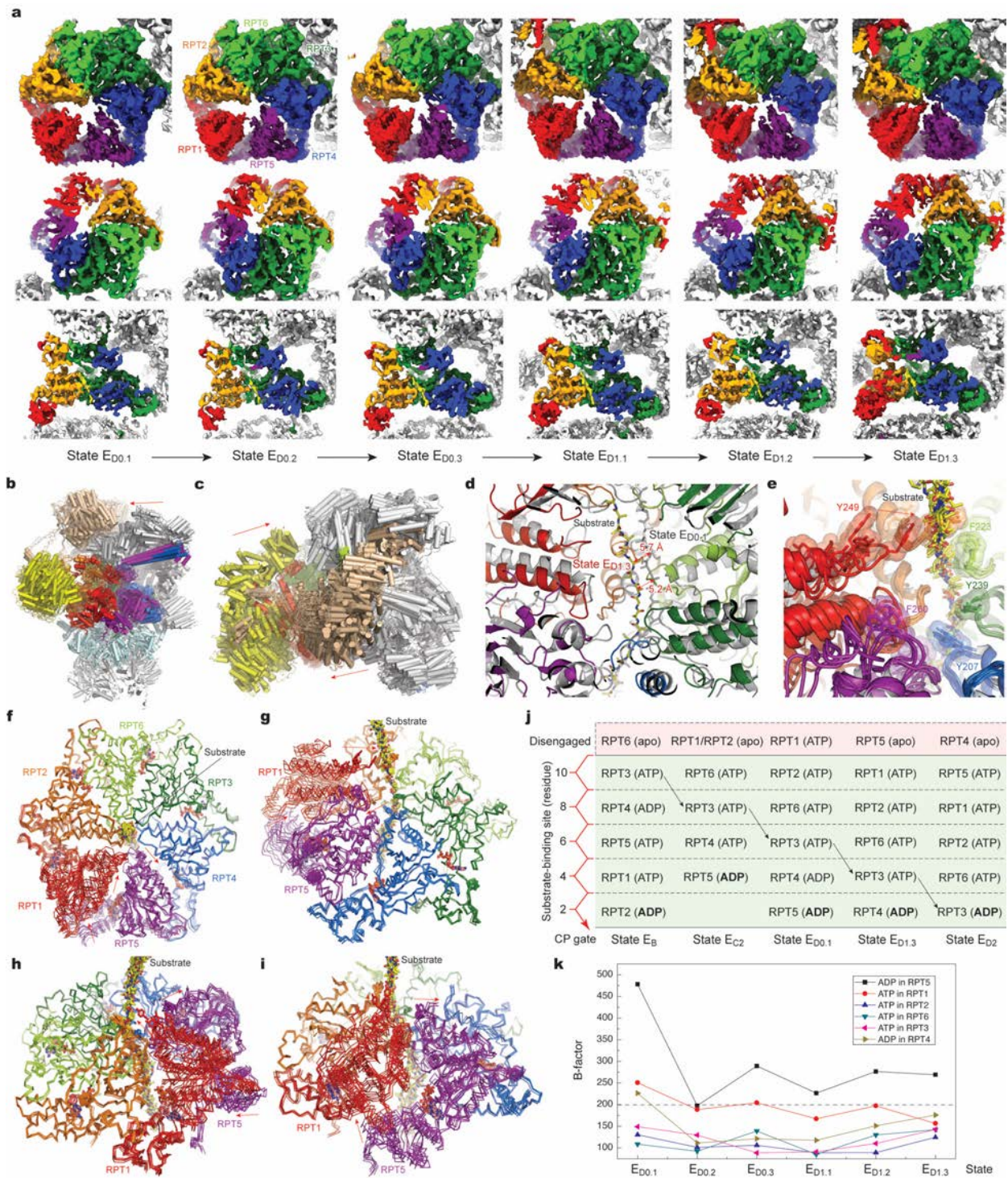

**Extended Data Fig. 10. Cryo-EM structures of six sequential intermediate states between $E_{C2}$ and $E_{D1.3}$.** **a**, Cryo-EM density maps of the AAA-ATPase motor in six sequential states from $E_{D0.1}$ to $E_{D1.3}$ from three different viewing angles. The substrate densities are highlighted in yellow. **b**, Superposition of the seven structures in cartoon representations from a lateral

perspective, with all structures aligned together based on their CP structures, showing the RP movements relative to the CP. **c**, Superposition of the six structures in cartoon representations from a top-view perspective, with all structures aligned together based on their CP structures. **d**, Measurement of substrate movement between states $E_{D0.1}$ and $E_{D1.3}$ relative to the CP, suggesting that the substrate is translated ~5-6 Å during the process of ADP release in RPT5 and of RPT1 re-association with the substrate. **e**, The closeup view of the pore-1 loop interaction with the substrate with all six states superimposed after aligned together based on the structures of RPT3, RPT4 and RPT6. **f-i**, Superposition of the six sequential states with their RPT3, RPT4 and RPT6 aligned together. Red arrows show the direction of subunit movements in RPT1 and RPT5. **j**, A diagram illustrating the axial stepping of the substrate-interacting pore-1 loops that is coupled with ATP hydrolysis in the RPT subunits, revised from the previously published results based on the present study[2]. **k**, The average B-factor of the nucleotides in six ATPase subunits fitted to the cryo-EM densities of states $E_{D0.1}$ to $E_{D1.3}$ and computed by Phenix in real-space refinement procedure[80]. The high B-factor of ADP fitted in RPT5 indicates its low occupancy or unstable association and suggests that the ADP in RPT5 undergoes dissociation in this process.

**Extended Data Table 1. Hyperparameters of the deep residual networks in the 3D autoencoder.** Optimizer: Adam. Epochs: 50. Initial learning rate: 0.01. A factor of 0.1 and patience of 3 means that the learning rate will times 0.1 (factor) if the loss function does not improve in 3 (patience) epochs. Batch size: 32 (for the three simulated datasets), 8 (for the substrate-bound proteasome dataset).

| Layer name | Output size (simulation) | Output size (proteasome) | Network structure | Number of kernels |
|---|---|---|---|---|
| Conv1 | $200 \times 200 \times 200$ | $300 \times 300 \times 300$ | $5 \times 5 \times 5$ | 2 |
| Conv2_x | $100 \times 100 \times 100$ | $150 \times 150 \times 150$ | $3 \times 3 \times 3$, stride 2 $3 \times 3 \times 3$ | 2 |
| Conv3_x | | | $3 \times 3 \times 3$ $3 \times 3 \times 3$ | 2 |
| Conv4_x | $50 \times 50 \times 50$ | $75 \times 75 \times 75$ | $3 \times 3 \times 3$, stride 2 $3 \times 3 \times 3$ | 1 |
| Conv5_x | | | $3 \times 3 \times 3$ $3 \times 3 \times 3$ | 1 |
| Conv6_x (encoding) | | | $3 \times 3 \times 3$ $3 \times 3 \times 3$ | 1 |
| TransConv7_x | $100 \times 100 \times 100$ | $150 \times 150 \times 150$ | $3 \times 3 \times 3$, stride 2 $3 \times 3 \times 3$ | 1 |
| TransConv8_x | | | $3 \times 3 \times 3$ $3 \times 3 \times 3$ | 1 |
| TransConv9_x | $200 \times 200 \times 200$ | $300 \times 300 \times 300$ | $3 \times 3 \times 3$, stride 2 $3 \times 3 \times 3$ | 2 |
| TransConv10_x | | | $3 \times 3 \times 3$ $3 \times 3 \times 3$ | 2 |
| TransConv11 | | | $5 \times 5 \times 5$ | 1 |

## Extended Data Table 2. Cryo-EM data collection, refinement and validation statistics.

|  | $E_{D0.1}$ | $E_{D0.2}$ | $E_{D0.3}$ | $E_{D1.1}$ | $E_{D1.2}$ | $E_{D1.3}$ | $E_{D2.1}$ |
|---|---|---|---|---|---|---|---|
| **Data collection and processing** | | | | | | | |
| Magnification | 105,000 | 105,000 | 105,000 | 105,000 | 105,000 | 105,000 | 105,000 |
| Voltage (kV) | 300 | 300 | 300 | 300 | 300 | 300 | 300 |
| Electron exposure (e–/Å$^2$) | 44 | 44 | 44 | 44 | 44 | 44 | 44 |
| Defocus range (μm) | -0.6 to -3.5 | -0.6 to -3.5 | -0.6 to -3.5 | -0.6 to -3.5 | -0.6 to -3.5 | -0.6 to -3.5 | -0.6 to -3.5 |
| Pixel size (Å) | 0.685 | 0.685 | 0.685 | 0.685 | 0.685 | 0.685 | 0.685 |
| Symmetry imposed | C1 | C1 | C1 | C1 | C1 | C1 | C1 |
| Initial particle images (no.) | 3,254,352 | 3,254,352 | 3,254,352 | 3,254,352 | 3,254,352 | 3,254,352 | 3,254,352 |
| Final particle images (no.) | 25,755 | 105,081 | 169,522 | 92,777 | 175,114 | 146,506 | 856,683 |
| Map resolution (Å) | 3.9 | 3.3 | 3.2 | 3.3 | 3.2 | 3.2 | 2.5 |
| FSC threshold | 0.143 | 0.143 | 0.143 | 0.143 | 0.143 | 0.143 | 0.143 |
| Map resolution range (Å) | 3.0-8.0 | 2.5-8.0 | 2.5-8.0 | 2.5-8.0 | 2.5-8.0 | 2.5-8.0 | 2.5-4.8 |
| **Refinement** | | | | | | | |
| Initial model used | | | | 6MSH | | 6MSJ | 6MSK |
| Model resolution (Å) | 4.3 | 4.1 | 4.0 | 3.9 | 3.8 | 3.9 | 3.2 |
|   FSC threshold | 0.5 | 0.5 | 0.5 | 0.5 | 0.5 | 0.5 | 0.5 |
| Model resolution range (Å) | 3.0-8.0 | 3.0-8.0 | 3.0-8.0 | 3.0-8.0 | 3.0-8.0 | 3.0-8.0 | 2.5-5.8 |
| Map sharpening $B$ factor (Å$^2$) | -20 | -20 | -20 | -20 | -30 | -30 | -20 |
| **Model composition** | | | | | | | |
| Non-hydrogen atoms | 106475 | 106475 | 106475 | 106475 | 106448 | 106448 | 106499 |
| Protein residues | 13540 | 13540 | 13540 | 13540 | 13540 | 13540 | 13540 |
| Ligands | 11 | 11 | 11 | 11 | 10 | 10 | 11 |
| $B$ factors (Å$^2$) | | | | | | | |
|   Protein | 155.88 | 155.03 | 155.53 | 126.38 | 112.34 | 133.02 | 88.54 |
|   Ligand | 268.45 | 173.9 | 201.79 | 134.06 | 115.01 | 144.41 | 88.86 |
| **R.m.s. deviations** | | | | | | | |
|   Bond lengths (Å) | 0.006 | 0.008 | 0.007 | 0.006 | 0.007 | 0.007 | 0.007 |
|   Bond angles (°) | 1.022 | 1.095 | 1.076 | 1.041 | 1.047 | 1.072 | 1.074 |
| **Validation** | | | | | | | |
| MolProbity score | 1.77 | 1.82 | 1.77 | 1.82 | 1.45 | 1.78 | 1.75 |
| Clashscore | 5.05 | 5.58 | 5.08 | 5.57 | 5.13 | 5.14 | 4.49 |
| Poor rotamers (%) | 0.29 | 0.47 | 0.39 | 0.44 | 0.35 | 0.36 | 0.37 |
| **Ramachandran plot** | | | | | | | |
| Favored (%) | 91.56 | 90.99 | 91.47 | 90.98 | 91.71 | 91.5 | 91.24 |
| Allowed (%) | 8.15 | 8.73 | 8.24 | 8.75 | 7.91 | 8.23 | 8.47 |
| Disallowed (%) | 0.29 | 0.28 | 0.28 | 0.27 | 0.39 | 0.28 | 0.29 |